\documentclass{SciPost}

\hypersetup{
    colorlinks,
    linkcolor={red!50!black},
    citecolor={blue!50!black},
    urlcolor={blue!80!black}
}

\usepackage[bitstream-charter]{mathdesign}
\DeclareSymbolFont{usualmathcal}{OMS}{cmsy}{m}{n}
\DeclareSymbolFontAlphabet{\mathcal}{usualmathcal}

\fancypagestyle{SPstyle}{
    \fancyhf{}
    \lhead{\colorbox{scipostblue}{\bf \color{white} ~SciPost Physics }}
    \rhead{{\bf \color{scipostdeepblue} ~Submission }}
    
    \fancyfoot[C]{\textbf{\thepage}}
}

\newcommand{\degrees}[0]{\ensuremath{^\circ}}

\usepackage{
    booktabs,
    subcaption
}

\begin{document}

\pagestyle{SPstyle}

\begin{center}
    {\Large \textbf{\color{scipostdeepblue}{AllShowers: One model for all calorimeter showers}}}
\end{center}
\begin{center}
    \textbf{
        Thorsten Buss\textsuperscript{1,2 \(\star\)},
        Henry Day-Hall\textsuperscript{2},\\
        Frank Gaede\textsuperscript{2},
        Gregor Kasieczka\textsuperscript{1} and
        Katja Krüger\textsuperscript{2}
    }
\end{center}

\begin{center}
    {\bf 1} Institut für Experimentalphysik, Universität Hamburg, Hamburg, Germany\\
    {\bf 2} Deutsches Elektronen-Synchrotron DESY, Hamburg, Germany
    \\[\baselineskip]
    \(\star\) \href{mailto:Thorsten.Buss@uni-hamburg.de}{\small Thorsten.Buss@uni-hamburg.de}
\end{center}

\section*{\color{scipostdeepblue}{Abstract}}
\textbf{%
    \boldmath{%
        Accurate and efficient detector simulation is essential for modern collider experiments. To reduce the high computational cost, various fast machine learning surrogate models have been proposed. Traditional surrogate models for calorimeter shower modeling train separate networks for each particle species, limiting scalability and reuse. We introduce AllShowers, a unified generative model that simulates calorimeter showers across multiple particle types using a single generative model. AllShowers is a continuous normalizing flow model with a Transformer architecture, enabling it to generate complex spatial and energy correlations in variable-length point cloud representations of showers. Trained on a diverse dataset of simulated showers in the highly granular ILD detector, the model demonstrates the ability to generate realistic showers for electrons, photons, and charged and neutral hadrons across a wide range of incident energies and angles without retraining. In addition to unifying shower generation for multiple particle types, AllShowers surpasses the fidelity of previous single-particle-type models for hadronic showers. Key innovations include the use of a layer embedding, allowing the model to learn all relevant calorimeter layer properties; a custom attention masking scheme to reduce computational demands and introduce a helpful inductive bias; and a shower- and layer-wise optimal transport mapping to improve training convergence and sample quality. AllShowers marks a significant step towards a universal model for calorimeter shower simulations in collider experiments.
    }
}

\vspace{\baselineskip}

\noindent\textcolor{white!90!black}{%
\fbox{\parbox{0.975\linewidth}{%
\textcolor{white!40!black}{\begin{tabular}{lr}%
  \begin{minipage}{0.6\textwidth}%
    {\small Copyright attribution to authors. \newline
    This work is a submission to SciPost Physics. \newline
    License information to appear upon publication. \newline
    Publication information to appear upon publication.}
  \end{minipage} & \begin{minipage}{0.4\textwidth}
    {\small Received Date \newline Accepted Date \newline Published Date}%
  \end{minipage}
\end{tabular}}
}}
}

\vspace{10pt}
\noindent\rule{\textwidth}{1pt}
\tableofcontents
\noindent\rule{\textwidth}{1pt}
\vspace{10pt}

\section{Introduction}\label{sec:intro}
Simulations of particle detectors in high-energy physics (HEP) experiments incur high computational costs, which are expected to increase beyond available resources in the near future\cite{HEPSoftwareFoundation:2017ggl,Boehnlein:2022}. Fast generative models must substitute the most expensive MC simulation steps to achieve sufficient statistics with the available computing resources. To fully realize the physics potential of new experiments with higher event rates and highly granular calorimeters, more accurate and efficient fast generative models must be developed.

Many techniques for fast calorimeter simulation have been explored for existing or similar to existing experiments; generative adversarial networks (GANs)~\cite{Paganini:2017hrr,Paganini:2017dwg,deOliveira:2017rwa,Erdmann:2018kuh,Erdmann:2018jxd,Musella:2018rdi,Belayneh:2019vyx,Butter:2020qhk,ATLAS:2020,Ghosh:2020kkt,ATLAS:2021pzo,ATLAS:2022jhk,FaucciGiannelli:2023fow,Dogru:2024gpk}, variational autoencoders (VAEs)~\cite{ATLAS:2022jhk,Cresswell:2022tof,Hoque:2023zjt,Liu:2024kvv}, classical normalizing flows (NFs)~\cite{Krause:2021ilc,Krause:2021wez,Schnake:2022,Krause:2022jna,Xu:2023xdc,Buckley:2023daw,Pang:2023wfx,Ernst:2023qvn,Schnake:2024mip,Du:2024gbp,Majerz:2025ykn}, auto-regressive models~\cite{Birk:2025wai}, and diffusion and continuous flow models~\cite{Mikuni:2022xry,Acosta:2023zik,Mikuni:2023tqg,Amram:2023onf,Jiang:2024ohg,Kobylianskii:2024ijw,Jiang:2024bwr,Favaro:2024rle,Brehmer:2024yqw}. Even the highly granular pixel vertex detector of Belle II has been simulated using a GAN~\cite{Hashemi:2023ruu}. Future calorimeter designs have also been specifically targeted in various generative modeling projects including; GANs~\cite{Carminati:2018khv}, VAEs~\cite{Diefenbacher:2023prl,Buhmann:2020pmy,Buhmann:2021lxj,Buhmann:2021caf,Buss:2025cyw,Bieringer:2022cbs}, NFs~\cite{Diefenbacher:2023vsw,Buss:2024orz} and continuous flow models~\cite{Buhmann:2023bwk,Buhmann:2023kdg,Buss:2025cyw,Raikwar:2025fky,Buss:2025kiu,Gaede:2025shc,Favaro:2026awn}.

These methods can be seen in comparison in a recent taxonomy of detector simulation~\cite{Hashemi:2023rgo}. For the simulation of current detector designs, the accuracies and efficiencies of many variants were compared in the CaloChallenge~2022~\cite{Krause:2024avx}.

The model put forward in this paper breaks new ground; to the best of our knowledge, no previous model has captured the response of both the electromagnetic calorimeter (ECAL) and hadronic calorimeter (HCAL) with such a comprehensive set of particle species, let alone for a highly granular Higgs factory detector.\footnote{\samepage{When preparing this manuscript for submission, \cite{Favaro:2026awn} was released. It attempts a similar task, albeit for a smaller set of different particles.}} This combination of twelve particle types in a single model would be challenging at current granularities, but is even more challenging with the high-granularity expected in future calorimeters.

A multi-particle model, such as this one, is needed for three reasons: firstly, in production environments, maintaining code infrastructure uses significant human and computational resources, and by handling twelve particles together, we simplify the codebase, reducing the cost of both validation and maintenance. Secondly, fast calorimeter simulation occurs within full-scale MC simulations, so the models must share local memory resources with many other components, which quickly becomes a limitation on model size and therefore performance. By combining particle types, common physics behavior will be shared between particles resulting in better use of local memory, and so facilitating more accurate modeling. Finally, it is hoped that fast calorimeter simulation might benefit a wider range of users than just the production environments at major experiments. These users could save energy and compute time on more dedicated small-scale tasks that might require custom installations. A comprehensive model for all particle showers reduces the technical overhead of setup and installation, increasing the utilization and usefulness of the model for these users.

To achieve this, we introduce AllShowers, a continuous normalizing flow (CNF) model with a Transformer architecture. AllShowers is trained on a diverse dataset of simulated showers in the highly granular calorimeters of the International Large Detector (ILD)~\cite{ILDConceptGroup:2020sfq}. The model consists of two components: the PointCountFM, which predicts the number of points per layer conditioned on incident particle information, and the CNF-transformer, which generates the position and energy of each point, additionally conditioned on the layer index of each point. AllShowers has several significant improvements compared to its predecessor models~\cite{Buss:2025cyw,Buss:2025kiu}. Using an embedding layer for the calorimeter layer index allows the model to learn all relevant calorimeter layer properties, such as material budget and distance from the calorimeter surface, from data. A custom attention masking scheme is employed to reduce computational demands and introduce a helpful inductive bias, allowing points to interact only with points in nearby layers. Additionally, a shower- and layer-wise optimal transport mapping is used to improve convergence during training and sample quality.

The layout of this paper is as follows. In the next section, section~\ref{sec:dataset}, the dataset is described. This includes a summary description of the detector chosen as an example of a detector system with high-granularity calorimeters, the particle gun used for shower generation, and the data preprocessing. Following this, in section~\ref{sec:model}, the architecture of the AllShowers model is presented, along with a description of the training process. Then the results are presented in section~\ref{sec:results}. Finally, in section~\ref{sec:conclusion}, the paper is concluded with a discussion of the findings.

\section{Dataset}\label{sec:dataset}
We used the International Large Detector (ILD)~\cite{ILDConceptGroup:2020sfq} as an example of a detector with highly granular sampling calorimeters. The ILD was initially designed for the International Linear Collider (ILC), a proposed electron-positron collider, and could be adapted for other future colliders. The ILD detector design is optimized for particle-flow algorithms, which reconstruct particles with high precision by combining information from multiple subdetectors.

The ILD calorimeter system consists of a highly granular electromagnetic calorimeter \linebreak[4] (ECAL)~\cite{CALICE:2008gxs} and a hadronic calorimeter (HCAL). Both of which sit within a superconducting coil generating a magnetic field of 3.5~T strength. The ECAL is composed of 30 layers with tungsten absorbers and silicon sensors with about \(5\times5\) mm\(^2\) cells. For mechanical reasons and to reduce dead material, two active layers are always mounted on either side of a tungsten support. This results in a small modulation in measured energy in even and odd layers. To improve energy resolution at low energies while preserving good confinement of most EM showers, two different absorber thicknesses are used: a smaller one for the first 20 layers and a larger one for the last 10 layers. The HCAL consists of 48 layers with stainless steel absorbers and polystyrene scintillator tiles measuring about \(3\times3\) cm\(^2\). The overall sampling fraction is on the order of 1\% in both calorimeters.

As in earlier work~\cite{Buss:2025kiu,Buss:2025bec}, we use a regularized readout geometry without insensitive gaps between calorimeter modules. This broadens the model's applicability to other incident point locations. Hits, which the model produces in inactive material, will be dropped when integrated into the full simulation chain.

Using Geant4~\cite{GEANT4:2002zbu,Allison:2006ve,Allison:2016lfl} and the DD4hep~\cite{Frank:2014} framework, we simulated a dataset of two million showers originating from twelve different incident particle types, namely: e\(^-\), e\(^+\), \(\pi^-\), \(\pi^+\), K\(^-\), K\(^+\), K\(^0_L\), p, \={p}, n, \={n}, and \(\gamma\). The incident particle type is randomly chosen for each shower with equal probability. We cover all incident angles allowing particles originating at the interaction point (IP) to reach the calorimeter barrel region, including magnetic-field effects. This means that the angular bounds depend on the incident particle's charge and energy. The energy of the incident particles is uniformly distributed between 5 GeV and 126 GeV. The simulation is started with a particle gun placed right in front of the calorimeter surface to prevent interactions within the tracing system. A random subset of 50k showers is used as a validation set. For testing, we simulated several datasets with the incident particle distributions and statistics given in the results section.

\subsection{Data Representation}
The calorimeter shower data are represented as a 4D point cloud of energy depositions (Geant4 steps) in active material, where each point is represented as a tuple \((x, y, z, e)\). Here, \(x\) and \(y\) denote the local coordinates in millimeters, with \(x\) aligned along the direction of the magnetic field, and \(z\) indicates the layer index, ranging from 0 to 77 (covering both ECAL and HCAL layers). \(e\) represents the deposited energy.

To reduce the number of points while preserving geometry independence, the energy depositions are binned into a grid that is three times finer in the two transversal dimensions, i.e., nine times higher granularity, than the respective cells~\cite{Buhmann:2023bwk}. For each non-empty bin, a point is created using the \(x\) and \(y\) coordinates of the highest energy deposition within the bin, and the total energy within the bin.

We counteract the incident angle dependence of the shower shape by shifting the \(x\) and \(y\) coordinates of each point such that the incident particle always appears to enter the calorimeter perpendicularly at the origin. While this transformation does not eliminate all angle dependencies, it significantly simplifies the model's learning task. The model still needs to learn all remaining angular dependencies, such as particles passing more material between the layers when entering under a steep angle. This is why it still needs to know the incident angle.

To further reduce the number of points, we remove all points with an energy deposit below 10 keV or with a time of over 200 ns (bunch crossing window). The time constriction is already applied before clustering. We place a quadratic bounding box around the shower core and remove all points outside this box. The side length is chosen to be the side length of the octagon formed by the ECAL surface (c.a. 1500 mm). This will exclude most of the points for which the flat layer assumption breaks. The excluded energy depositions are far away from the shower core and typically low in energy. After these preprocessing steps, the average number of points per shower is 2306, with a maximum of 6006.

For preprocessing, the \(x\) and \(y\) coordinates are rescaled to have standard deviation one and mean zero (Standardization), \(z\) is kept as the discrete layer index, and the logarithm of the energy is also standardized. Points are zero-padded to a maximum of 6016 points per shower for batch training. 6016 is a multiple of 128, making the computation of attention masks easier and more efficient.

\section{Model and Training}\label{sec:model}
\begin{figure}
    \centering
    \includegraphics{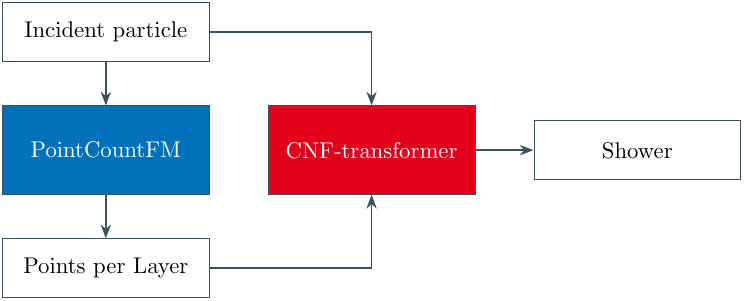}
    \caption{Schematic overview of the AllShowers model architecture. The PointCountFM predicts the number of points per layer, which are then used by the CNF-transformer to generate the full shower. The incident particle information is provided to both models. The point layer index, needed by the CNF-transformer, can be computed from the number of points per layer.}
    \label{fig:model}
\end{figure}
The AllShowers model consists of two main components: the PointCountFM and the CNF-transformer, as illustrated in figure~\ref{fig:model}. The PointCountFM is responsible for generating the number of points per layer conditioned on the incident particle information (particle type, energy and angle). After initializing as many points as demanded by PointCountFM, the CNF-transformer generates the position within the layer and the energy of each point, again conditioned on the incident particle information. The layer index is provided as an additional condition. Splitting the model into two components in this way is inspired by CaloFlow~\cite{Krause:2021ilc}, where the layer-wise energy depositions are generated first, and has been used in various other works.

Both models use the continuous normalizing flow (CNF)~\cite{Chen:2018} paradigm as a means to model the complex distributions of calorimeter showers. In CNFs, the transformation from latent to physics space is modeled as the solution of an ordinary differential equation (ODE) \(\frac{d\mathbf{x}_t}{dt}=v_c(x_t,t)\), where \(v_c(\mathbf{x}_t,t)\) is a neural network that predicts the vector field, \(t\) is the integration variable, and \(c\) is the condition. \(\mathbf{x}_0\) is the initial state, a sample from the latent space. \(\mathbf{x}_1\) is a physics space sample. Note that \(x\) denotes the spatial coordinate in the calorimeter while \(\mathbf{x}_t\) denotes an entire data sample. During generation, a numerical ODE solver is used.

A likelihood-based training of CNFs is possible~\cite{Chen:2018}, but computationally inefficient. Instead, we use the recently proposed conditional flow matching (FM)~\cite{Lipman:2023} approach. In FM, the vector field is constructed as the expectation value of all straight lines connecting physics and latent space samples. This mean squared error is evaluated using Monte Carlo integration over physics and latent space. FM has been shown to be more efficient than likelihood-based training for CNFs~\cite{Lipman:2023}.

\subsection{PointCountFM}
The PointCountFM was already introduced in CaloHadronic~\cite{Buss:2025cyw}. It is responsible for generating the number of points per layer, 78 integers in total, conditioned on the incident particle information (type, energy, angle). The type is given as a one-hot encoded vector, the energy is converted to logarithmic scale and standard scaled, and the angle is represented by a vector on the unit sphere. This is a generalization of the approach taken in CaloHadronic, where only fixed angle and particle type were considered.

As an additional improvement, we no longer use dequantization noise during training. While dequantization is essential for classical likelihood-based training of flows on discrete data, it is not necessary for FM. We found that removing the dequantization noise leads to a significant improvement in performance, especially for low point counts. Dequantization works well when a change by one in the discrete value has no significant effect on the downstream task. However, in our case, a change from zero to one point in a layer can confuse the CNF-transformer significantly, as it has to generate a point in an unexpected layer.

A complete list of hyperparameters can be found in Appendix~\ref{app:hyperparameters}.

\subsection{CNF-Transformer}
\begin{figure}
    \centering
    \includegraphics{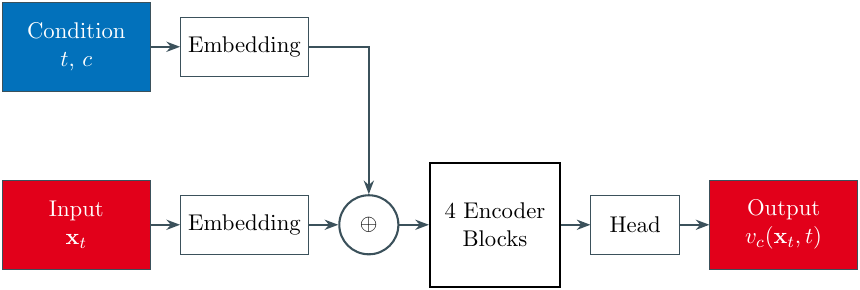}
    \caption{Schematic overview of the CNF-transformer architecture. The input \(\mathbf{x}_t\) for \(t=0\) is a standard normal sample, for \(t=1\) it is the preprocessed shower. Since the calorimeter layer is a condition, \(\mathbf{x}_t\) is a three-dimensional point-cloud \((x_k, y_k, e_k)\). The condition \(c\) includes the incident particle information and the layer index. \(t\) is the time variable of the neural ODE. The output \(v_c(\mathbf{x}_t, t)\) is the vector field used in the CNF, e.g. the right-hand side of the neural ODE.}
    \label{fig:cnf-transformer}
\end{figure}
After PointCountFM has predicted the number of points per layer (\(n_i\)), as many latent space points as requested are initialized. The first \(n_0\) points are assigned to layer 0, the next \(n_1\) points to layer 1, and so on. Each point is initialized with a standard normal sample in the \(x\), \(y\), and \(\log(e)\) dimensions. The layer index, \(z\), is provided as an additional condition. Then CNF-transformer transforms these points into a calorimeter shower.

An overview of the CNF-transformer architecture is shown in figure~\ref{fig:cnf-transformer}. The input is the point cloud, \(\mathbf{x}_t\), at ODE time \(t\). For \(t=0\), this is a standard normal sample, and for \(t=1\), it is the preprocessed shower. The output is the vector field \(v_c(\mathbf{x}_t, t)\) used in the CNF, i.e. the right-hand side of the neural ODE. We can split the condition into global conditioning information and point-wise conditioning information. The global conditioning information includes the incident particle type, energy, and angle, while the point-wise conditioning information is the layer index. Input, time, and conditions are embedded and element-wise summed. The resulting representation is processed by four transformer encoder blocks. Finally, a head network produces the output vector field.

\subsection{Embeddings}\label{sec:embeddings}
The main purpose of the embeddings is to map the different inputs to a common feature space.

\paragraph{Input Embedding}
The input \(\mathbf{x}_t\) is a point cloud of shape \((N, 3)\), where \(N\) is the number of points. We embedded each point independently using a single linear layer going from 3 to 64 dimensions.

\paragraph{Time Embedding}
For the time embedding, we used the standard Fourier feature mapping~\cite{Tancik:2020} with 3 frequencies, followed by a linear layer going from 6 to 64 dimensions.

\paragraph{Condition Embedding}
\begin{figure}
    \centering
    \includegraphics[width=0.65\textwidth]{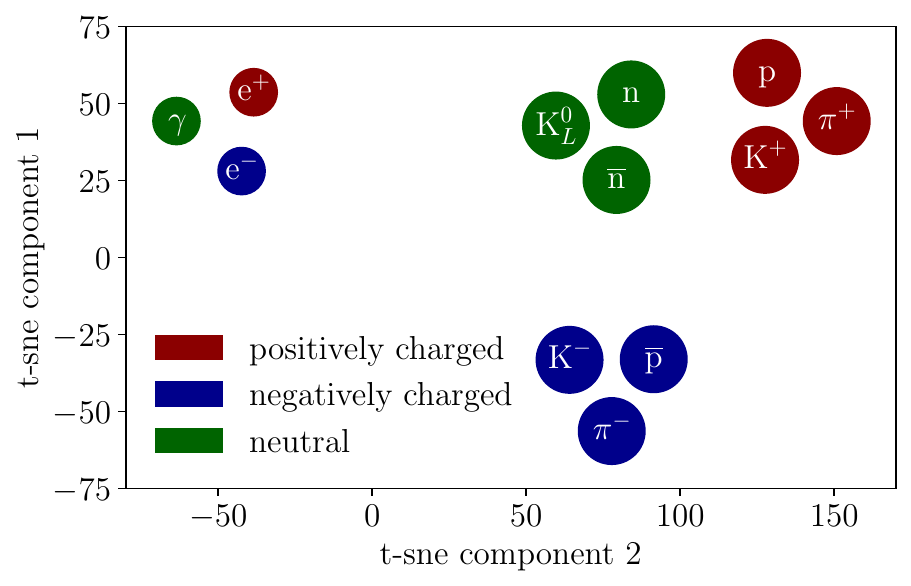}
    \caption{2D t-SNE~\cite{Maaten:2008} visualization of the twelve learned 64-dimensional particle type embeddings. Small circles indicate electromagnetic showers, large circles indicate hadronic showers. One can see four distinct clusters: one for electromagnetic showers, one for positively charged hadrons, one for negatively charged hadrons, and one for neutral hadrons.}
    \label{fig:tsne_particle_embeddings}
\end{figure}
The global conditional information includes the incident particle type, energy, and direction. The model explicitly learns twelve embedding vectors, one for each particle type. Figure~\ref{fig:tsne_particle_embeddings} shows a t-SNE~\cite{Maaten:2008} visualization of these embeddings. The preprocessed energy and direction are concatenated and passed through a linear layer going from 4 to 64 dimensions. Since the point layer is provided as point-wise information, the model has an implicit conditioning on the number of points per layer. To make it explicit, we also provide the number of points per layer as global information. The number of points per layer is passed through a feedforward network with one hidden layer of size 128 with ReLU activation, going from 78 to 64 dimensions.

\paragraph{Layer Embedding}
The calorimeter layer index is provided as point-wise conditional input. The model explicitly learns 78 embedding vectors, one for each of the 78 layers. This allows the model to learn layer-specific features like distance from the ECAL surface, material budgets, and typical energy deposition.
\\[\baselineskip]
After embedding, the global features are repeated for each point and all features are summed element-wise.

\subsection{Fast Attention Masking}
\begin{figure}
    \centering
    \includegraphics[width=0.3\textwidth]{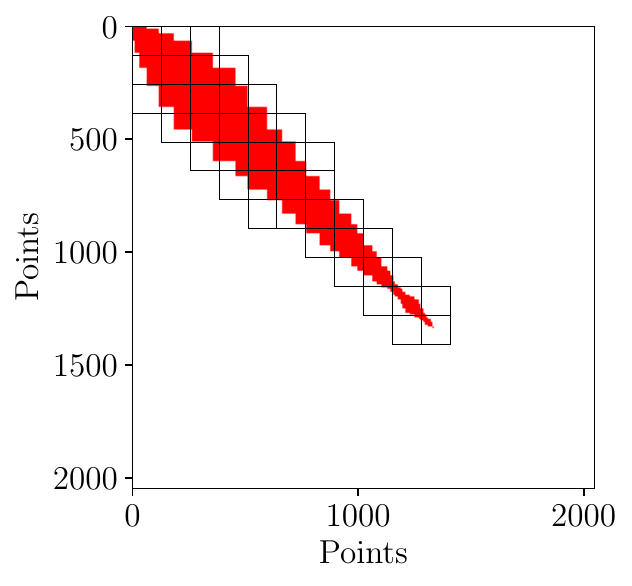}
    \includegraphics[width=0.3\textwidth]{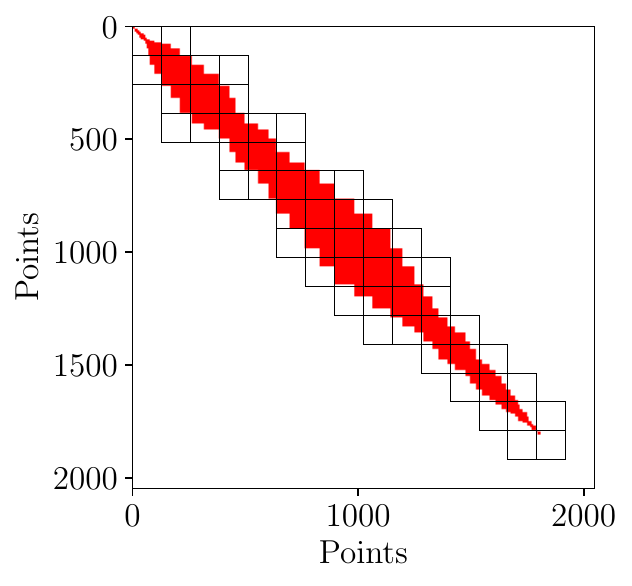}
    \includegraphics[width=0.3\textwidth]{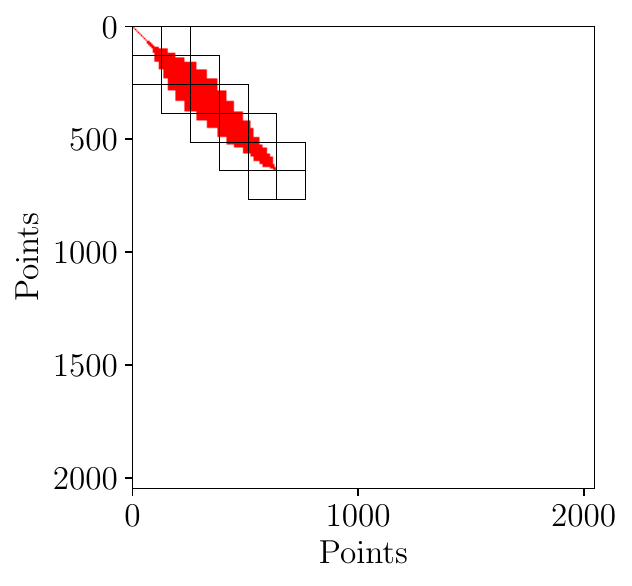}
    \caption{Examples of attention masks for three different showers. Shown is a part of the full \(6016 \times 6016\) attention matrix where each entry indicates whether two points can interact with each other. Red entries indicate allowed attention, white entries indicate masked attention. The black lines indicate the \(128 \times 128\) blocks flex-attention~\cite{Dong:2024} will compute. White entries within these blocks are computed but then masked out.}
    \label{fig:attention_masks}
\end{figure}
One major drawback of transformers is their quadratic complexity in the number of input tokens. In our dataset, the number of points per shower can be up to 6016, which would require more than 36 million attention weights. In the computer science literature, various methods for masking attention weights have been proposed. Most notably, the Sparse Transformer~\cite{Child:2019}, Longformer~\cite{Beltagy:2020}, and BigBird~\cite{Zaheer:2020} architectures. However, these methods have been developed in the context of natural language processing, where the input is a sequence. In our case, the input is a point cloud without any inherent ordering.

We developed a custom attention masking scheme that takes advantage of the fact that points are grouped by calorimeter layer. We allow points that are up to two layers apart to interact with each other. This means that points in layer \(i\) can interact with points in layers \(i-2\), \(i-1\), \(i\), \(i+1\), and \(i+2\). In combination with padding masking~\cite{Vaswani:2017}, this leads to a high degree of sparsity in the attention weights for our dataset. To utilize this sparsity, we used PyTorch's~\cite{PyTorch} built-in FlexAttention~\cite{Dong:2024} module. For this to work efficiently, input points are sorted by layer index before starting the training. An example of attention masks for three different showers is shown in figure~\ref{fig:attention_masks}.

This attention masking scheme leads to a significant speed-up during training and inference, roughly a factor of 20, which is consistent with the sparsity of the attention weights. This allowed us to train the CNF-transformer for more epochs, leading to better performance. We also found improved performance with the same number of training epochs, indicating that the inductive bias introduced by the attention masking is beneficial.

\subsection{Layer-wise Optimal Transport Mapping}
In continuous normalizing flows and diffusion models, the sampling process involves transforming samples from a simple latent distribution (e.g., a standard normal distribution) to match the complex data distribution. To do so, an ordinary or stochastic differential equation (ODE or SDE) is solved, which can be resource-intensive. The number of function evaluations (NFE) necessary to get good results is strongly correlated with the curvature of the trajectories taken by samples during the transformation \(\kappa = \left|\ddot{\mathbf{x}}_t\right|\), where \(\ddot{\mathbf{x}}_t\) is the second derivative of the sample with respect to the integration variable \(t\).

The main reason CNFs have curvature in their trajectories is the random mapping of data points to latent points during training. To overcome this problem, batch-wise optimal transport (OT) mapping has been proposed~\cite{Tong:2024}. The idea is to approximate the optimal mapping between data points and latent points, which would lead to straight trajectories. To achieve this, the optimal transport problem is solved for each batch during training. However, this approach is only feasible for generative problems without or with simple conditioning.

Instead of mapping data and latent points, we map physics point-cloud points to latent point-cloud points, exploiting the permutation invariance. Since the calorimeter layer conditioning breaks permutation invariance, the OT mapping is only applied per shower and layer. We solve the OT problem using the Python Optimal Transport (POT) library~\cite{POT}. The cost function is the Euclidean distance in the 3D space of preprocessed points.

The layer-wise OT mapping leads to shorter trajectories, faster training convergence, and better results.

\subsection{Training Details}
We trained the CNF-transformer using the Lookahead optimizer~\cite{Zhang:2019} with RAdam~\cite{Kingma:2014,Liu:2021} as the inner optimizer and decoupled weight decay~\cite{Loshchilov:2017}. We found this combination, also known as Ranger~\cite{Wright:2019}, to be especially robust against training instabilities, leading to reliable convergence in our experiments. We wrote a custom Ranger implementation in PyTorch to fit our needs. As learning rate scheduler, we used a cosine annealing schedule. Since RAdam has an integrated warm-up phase, we did not use an additional warm-up schedule. We trained the model with a batch size of 256 for 200 epochs. The training took less than 24 hours on 16 Nvidia A100 GPUs. A complete list of hyperparameters can be found in Appendix~\ref{app:hyperparameters}.

\subsection{Energy Calibration}\label{sec:energy_cali}
After training, we found that the total energy per shower generated by the CNF-transformer was too low by approximately \(3.3\%\) on average. As the total energy is not directly modeled, but rather indirect as the sum of the point energies and it is strongly correlated with the incident energy, it is difficult for the model to learn to get this exactly right. Because this is a particularly important quantity in calorimeter simulation, we decided to apply a simple calibration step. While a simple rescaling of the point energies could fix this, it would influence other distributions in a negative way. Instead, we rescale the incident energy we provide as condition to the CNF-transformer by a factor of \(1.033\) during inference. This simple calibration step fixes the total energy per shower without negatively impacting other distributions.

\section{Results}\label{sec:results}
In the following section, the performance of the model is presented on multiple levels.
Beginning at the single event scale, in section~\ref{sec:individual_showers}, the ability of the model to generate realistic, detailed events is shown.
Secondly, in section~\ref{sec:distributions}, the ensemble-level distributions of the model are compared to the targets they seek to replicate.
Following this, in sections~\ref{sec:comparison_calo_clouds} and~\ref{sec:comparison_calo_hadronic}, the ensemble-level distributions are compared to other models with similar objectives,
and finally, section~\ref{sec:timing} looks at the inference speed of this model.

In order to render all comparisons fair, the same post-processing is applied to the output of all models.
Hits produced by the models are clustered into regular grids intended to resemble the granularity of the calorimeter in question;
so in the ECAL, hits have been clustered into cells of \(5 \times 5\) millimeters, and in the HCAL, into cells of \(30 \times 30\) millimeters.
Each model has used its own conventions for training data preprocessing, and we do not wish the relationship between the
grid in post-processing and any grids imposed on the training data to introduce artifacts. Therefore, we add a random offset to the post-processing grid in each event.
Finally, cells with energy below half the energy deposited by a Minimum Ionizing Particle (MIP), 0.1 MeV in the ECAL and 0.25 MeV in the HCAL, are conventionally removed before reconstruction to reduce electronics noise. Even though there is no electronics noise in our simulated data, we remove these cells during post-processing to be closer to an application in a production environment.

\subsection{Individual Showers}\label{sec:individual_showers}
\begin{figure}
    \centering
    \includegraphics[width=0.295\textwidth]{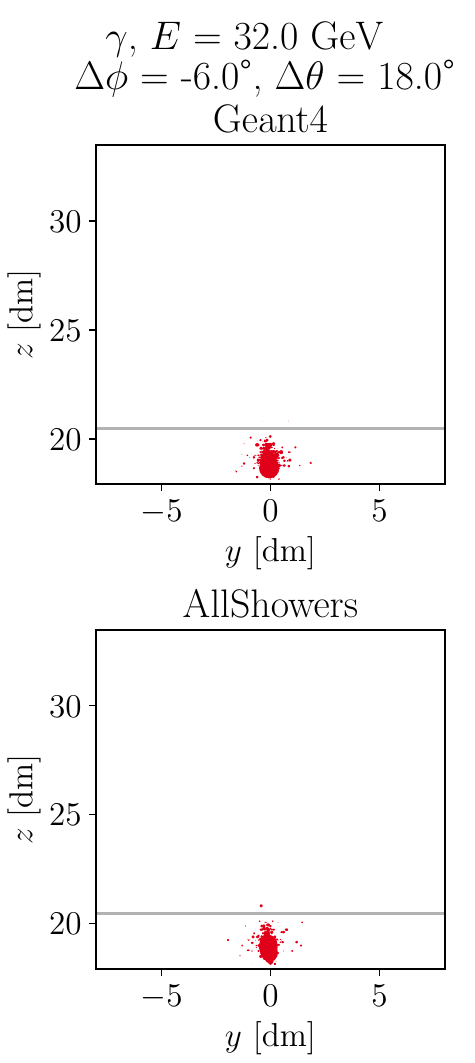}
    \includegraphics[width=0.295\textwidth]{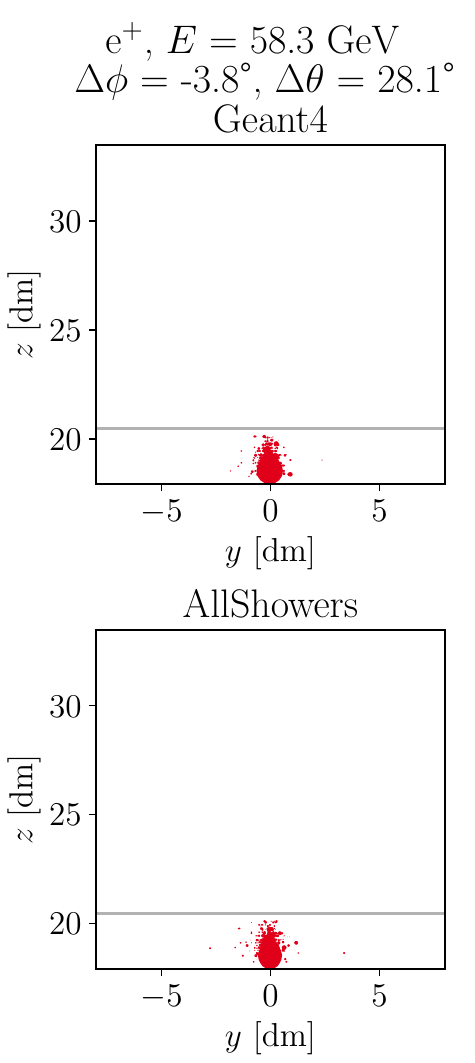}
    \includegraphics[width=0.295\textwidth]{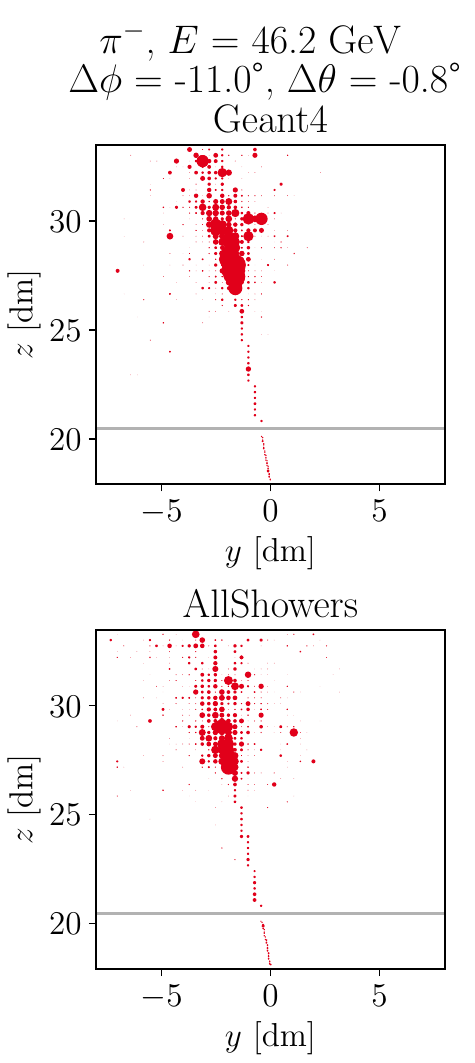}

    \vspace{1.5em}

    \includegraphics[width=0.295\textwidth]{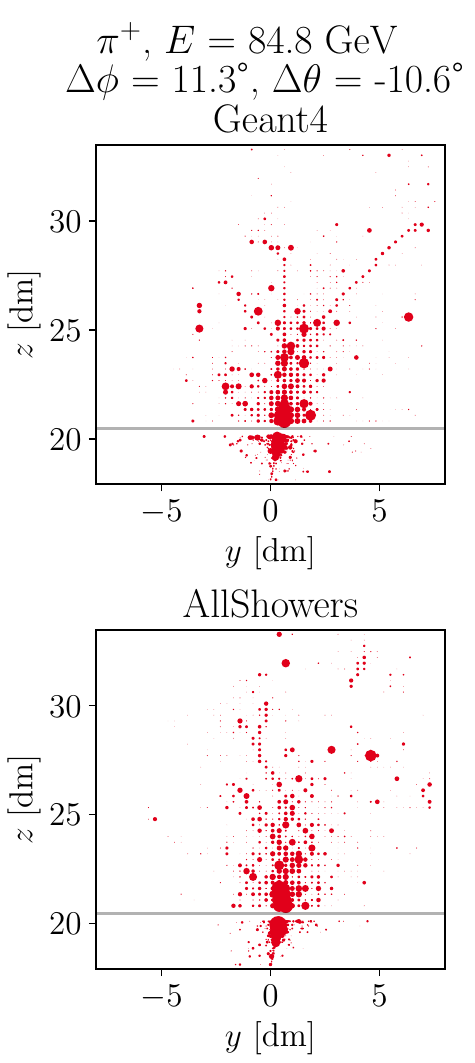}
    \includegraphics[width=0.295\textwidth]{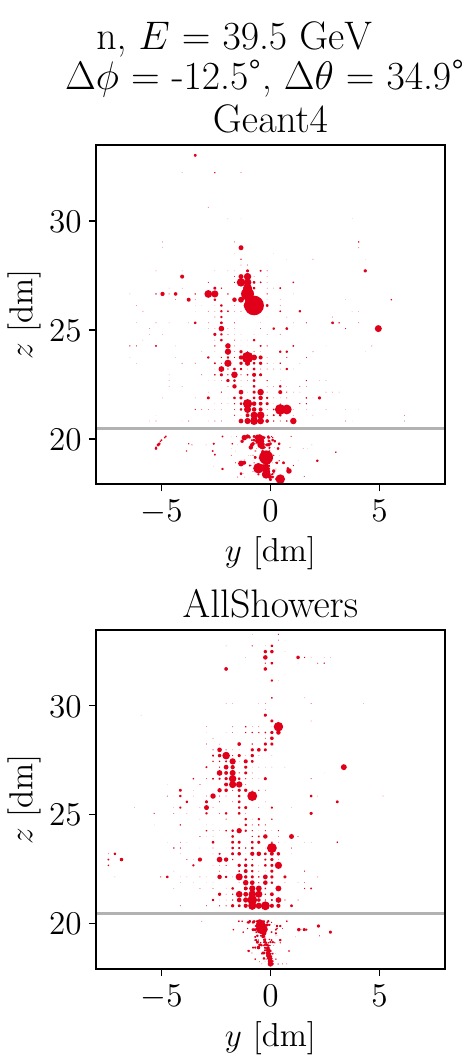}
    \includegraphics[width=0.295\textwidth]{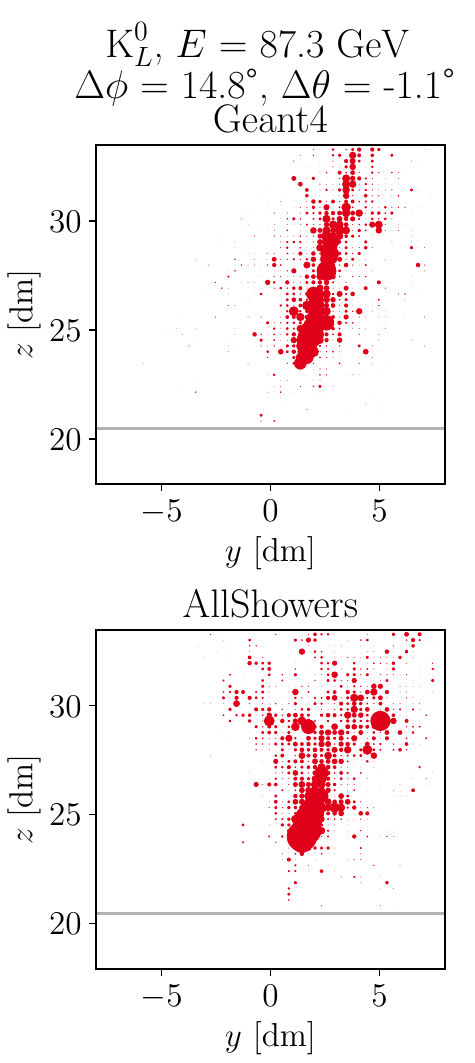}

    \caption{Comparison of individual showers simulated with Geant4 and with AllShowers for different incident particles, energies and angles. The point size indicates the energy of each hit. For these showers, the number of points per layer was taken from the Geant4 simulation rather than generated by the PointCountFM to allow for a more direct comparison of the spatial and energy distribution of hits.} 
    \label{fig:individual_showers}
\end{figure}
One of the more exciting features of a high-granularity calorimeter is how distinctly it resolves particle showers from different particle types.
In figure~\ref{fig:individual_showers}, we can see examples of six particle types, each shown once as simulated by Geant4 (upper) and once by AllShowers (lower).
The direction and energy chosen are the same for each model, and the number of points per layer is fixed to be that chosen by the Geant4 simulation.
That is to say, for AllShowers, only the CNF-transformer is used; PointCountFM does not run.
Thus, the two models are compelled to generate events with similar depth for each shower, and the results are directly comparable.
In each image, a gap can be seen at about \(z=2015\)~mm where the ECAL ends and the HCAL has yet to start.

It is clear from these images that the embeddings (see section~\ref{sec:embeddings}) used to encode the particle type are sufficient for AllShowers to produce appropriately tailored behavior in each shower.
By eye, not only is each particle type distinct, but the features align well with those seen in the Geant4 simulation.
The electromagnetic showers of the \(\gamma\) and \(e^+\) each have the typical cloud-like distribution, well contained in the ECAL.
The charged pions (\(\pi^+\) and \(\pi^-\)) each have well-defined MIP tracks in AllShowers, which correctly point to the start of the shower.
For the \(\pi^-\), this requires traversing right through the ECAL into the HCAL.
Each pion shows a mild bend of the MIP track in opposite directions to account for the response to the magnetic field.
Both pions then shower, with AllShowers displaying marginally fewer defined secondary tracks than Geant4, but still providing some, and displaying a very plausible shower pattern.
The neutron (\(n\)) event produced by AllShowers also replicates the overall shower cone well, again perhaps showing fewer secondaries.
Finally, as appropriate for a neutral particle, AllShowers does not generate a MIP track for the \(\mathrm{K}^0_L\) particle. The feature of neutral hadrons is hidden for the neutron shower shown here, since it starts showering immediately upon entering the calorimeter.
The \(\mathrm{K}^0_L\) shower develops in AllShowers with good substructure, including visible internal secondary tracks, and a correct funnel shape.
The shower start is marginally less aggressive in AllShowers than in Geant4, but it is a very plausible \(\mathrm{K}^0_L\) shower.

\subsection{Distributions}\label{sec:distributions}
\begin{figure}
    \centering
    \includegraphics[width=0.96\textwidth]{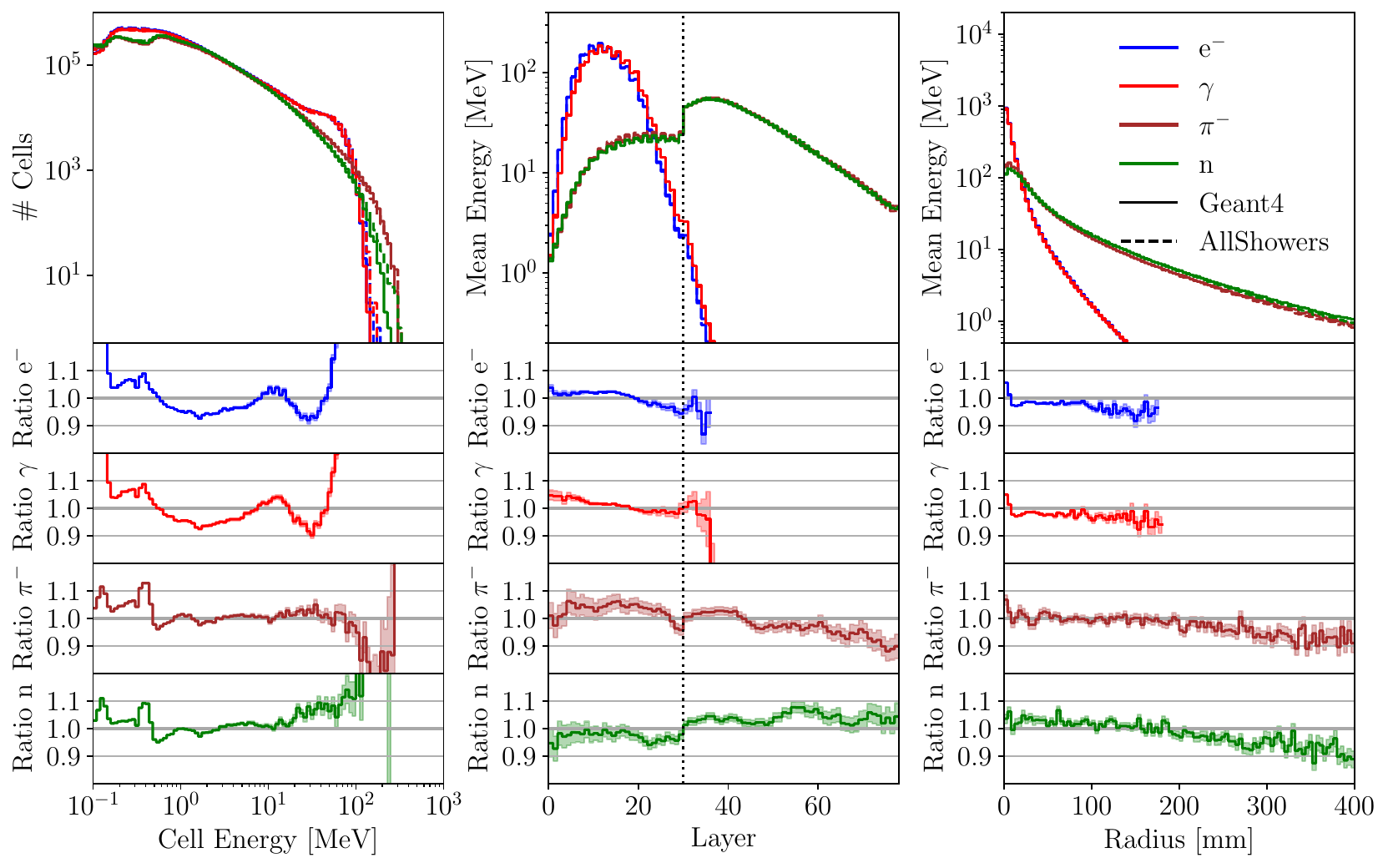}
    \includegraphics[width=0.96\textwidth]{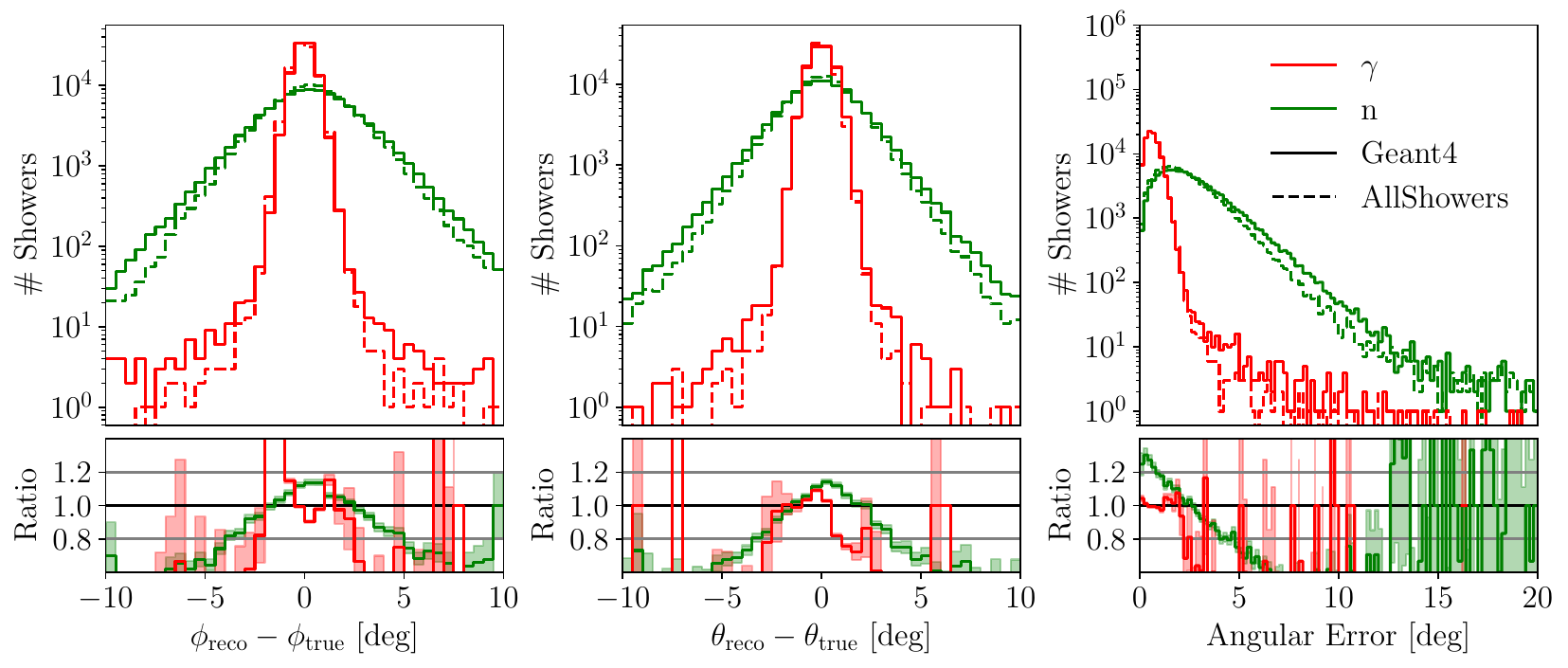}
    \caption{Histograms comparing Geant4 and AllShowers for different incident particle types and angles. The incident energy is fixed at 100 GeV. Top row from left to right: cell energy spectrum, longitudinal energy distribution, and radial energy distribution. Bottom row plots show principle component based angular reconstruction errors. The solid lines represent Geant4 and the dashed lines AllShowers. The lower panels of each plot show the ratio of AllShowers over Geant4. For the longitudinal and radial energy distributions, the standard deviation of the mean and for all other distributions the Poisson error is shown as error bars. Per particle type and generator, 10k showers were simulated/generated for the upper row and 100k samples for the lower row.}
    \label{fig:distributions_all_types}
\end{figure}

While having visually credible individual showers is clearly an asset, almost all physics analysis happens on the ensemble-level.
In figure~\ref{fig:distributions_all_types}, we present histograms comparing kinematic behavior of Geant4 and AllShowers for selected particles; \(e^-\), \(\gamma\), \(\pi^-\) and \(n\). Other particles have similar accuracy, but are omitted for clarity in the plots. The omitted particles can be found in \autoref{app:additional_plots}.

In the top panels, both AllShowers (dashed line) and Geant4 (solid line) are shown for each of three quantities: cell energy spectrum, longitudinal energy distribution and radial energy distribution. For most values, the top panels show no observable difference between Geant4 and AllShowers. In the lower panels, ratios of AllShowers to Geant4 are shown.

The most striking aspect of these plots is the clear dimorphism of electromagnetic showers (\(e^-\) and \(\gamma\)) and hadronic showers (\(\pi^-\) and \(n\)).
This bimodal behavior is well known, and that AllShowers accurately captures both variants demonstrates its flexibility.

In the cell energy spectrum on the left, AllShowers is within \(10\%\) of Geant4 for most of the range in all particle types.
It makes a good replication of the MIP peak near \(10^{-1}\) GeV, and does not significantly deviate until we reach the sparsely populated tails of the spectrum.
The high-energy tails tend to be somewhat overpopulated in AllShowers; the poor modeling is likely due to scarcity of this region in the training data.

In the longitudinal energy distribution in the center, the same dimorphism between electromagnetic and hadronic showers is clear.
AllShowers's behavior here is strongly influenced by the performance of the PointCountFM, and the agreement with Geant4 is within \(10\%\) for all but the extreme tails.
This plot emphasizes the value of also modeling the HCAL for electromagnetic showers: \(\gamma\) showers in particular are not always well contained to the ECAL, and AllShowers manages to capture the tail that bleeds into the HCAL.

Finally, we see the dimorphism again in the radial distribution on the right.
This radial distribution shows remarkably good agreement for the bulk of the shower.
At the innermost core, some deviation is visible, but still within \(10\%\) of Geant4 for all particles.
While there is more deviation in the tails, there are very few particles in these regions to work with, so it is expected that model performance may not be optimal here.

The lower row shows results of internal angle reconstructions based on a simple principal component analysis (PCA) with outlier removal. This way of internal angle reconstruction only works for electrically neutral particles where we do not have tracking information and the initial particle is not effected by the magnetic field. We see that AllShowers reproduces the Geant4 distribution well overall. The difference between photons and neutral hadrons was well learned by the model. Compared with the profile plots, we observe somewhat larger deviations. This is expected since the model does not learn this quantity directly but only as a corelation in its input space which is known to be more challenging for generative models. More details on the angle reconstruction can be found in \autoref{app:angular_reconstruction}.

\subsection{Comparison to CaloClouds3}\label{sec:comparison_calo_clouds}
For the case of photons only, we can compare the performance to the performance of the CaloClouds3 model~\cite{Buss:2025kiu}. CaloClouds3 is a fast generative diffusion model, specialized to only photon showers, trained on the ECAL only. As current generative models would not be applied in regions where different layer orientations meet, we also restrict the comparison data to photon showers, with \(45\degrees < \theta < 135\degrees\) and \(79\degrees < \phi < 109\degrees\). An energy range is chosen such that it sits comfortably inside both models' training regions; \(10\) to \(90\) GeV.

\begin{figure}
    \centering
    \includegraphics[width=0.97\textwidth]{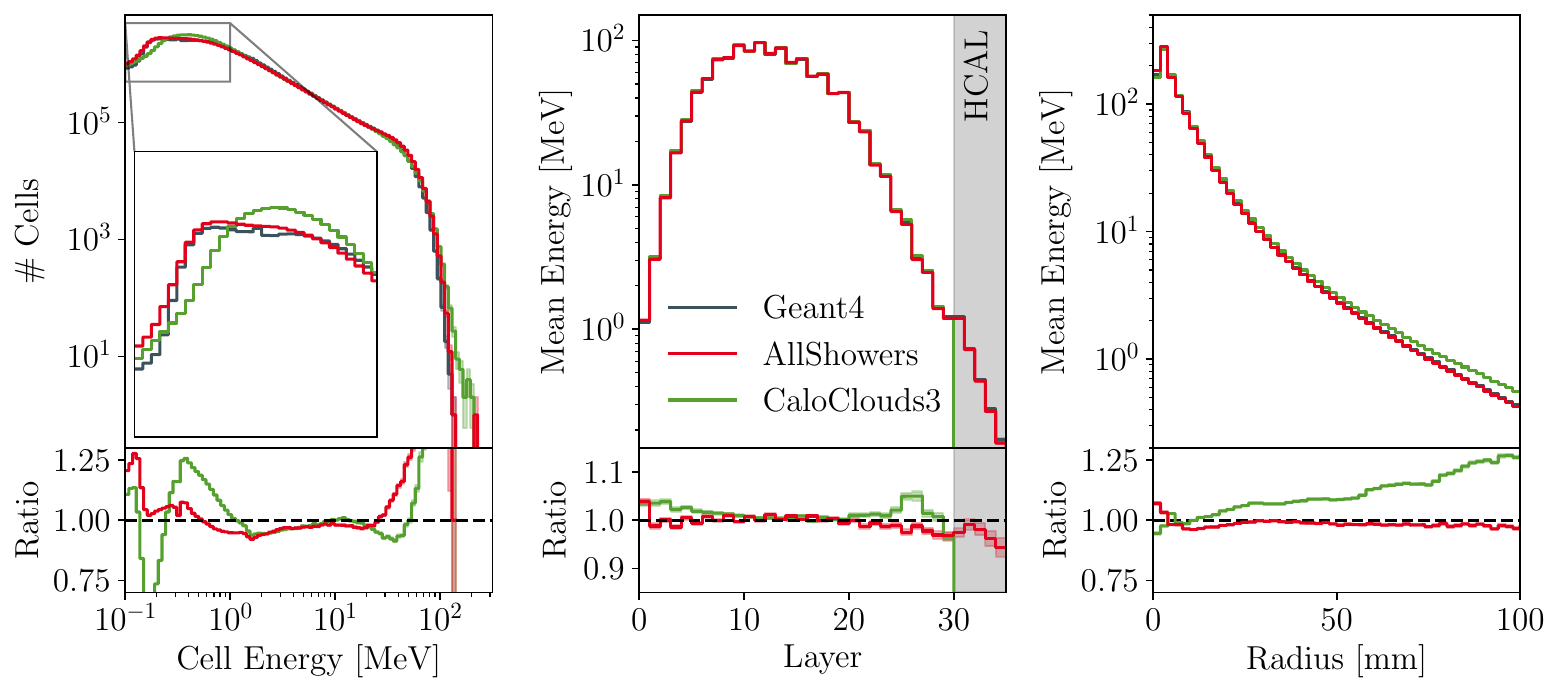}
    \caption{Comparison of AllShowers and CaloClouds3 on photon showers with incident energies uniformly distributed between 10 GeV and 90 GeV. Incident angles are distributed over the intersection of the respective training regions. From left to right: cell energy spectrum, longitudinal energy distribution and radial energy distribution. Per generator, 50k samples are used.}
    \label{fig:comparison_calo_clouds}
\end{figure}

In figure~\ref{fig:comparison_calo_clouds}, the standard three kinematic profiles are shown for both fast models and Geant4. On the left, the cell energy spectrum for AllShowers is notably better aligned with Geant4 than CaloClouds3; in particular, AllShowers has a well-formed replication of the MIP peak near \(10^{-1}\) GeV. Neither model quite fits the high-energy tail, but with very few data points, this is a challenging region to learn.

In the center, CaloClouds3 and AllShowers perform equally well on the longitudinal energy distribution. CaloClouds3 is a little better at replicating the alternating layer pattern, but tends to overpopulate the start and end of the shower. On the right-hand side of this plot, a grey band indicates the HCAL, for which only AllShowers has training data. This region is about as populated as the smallest bin in the ECAL, so its contribution is not negligible, and AllShowers's capacity to capture this information would be valuable in advanced reconstructions.

Finally, in the radial distribution on the right, AllShowers is significantly better than CaloClouds3. It maintains a flat ratio to Geant4 right out into a long distribution tail, and only marginally misrepresents the center of the shower. CaloClouds3 is unable to keep a flat ratio and deviates significantly from Geant4 towards the tail. While the deviation of CaloClouds3 in the tail here seems very large, it should be noted that the comparison of machine learning models ultimately will have to be done on physics observables, computed after a full event reconstruction has been applied, and that, as shown in~\cite{Buss:2025bec}, the CaloClouds3 model performs reasonably well on \(\pi^0\)-reconstruction.

Conceptually, the key distinction between AllShowers and CaloClouds3 is that the diffusion model in CaloClouds3 generates points which are independent and identically distributed (iid). There are longitudinal correlations, imposed by the normalizing flow component of CaloClouds3, but these are not known to the diffusion model, and they only describe the macro features, energy per layer and points per layer. This means that CaloClouds3 cannot capture point-to-point correlations. By contrast, AllShowers entertains correlations between points themselves.

\begin{figure}
    \centering
    \includegraphics[width=0.95\textwidth]{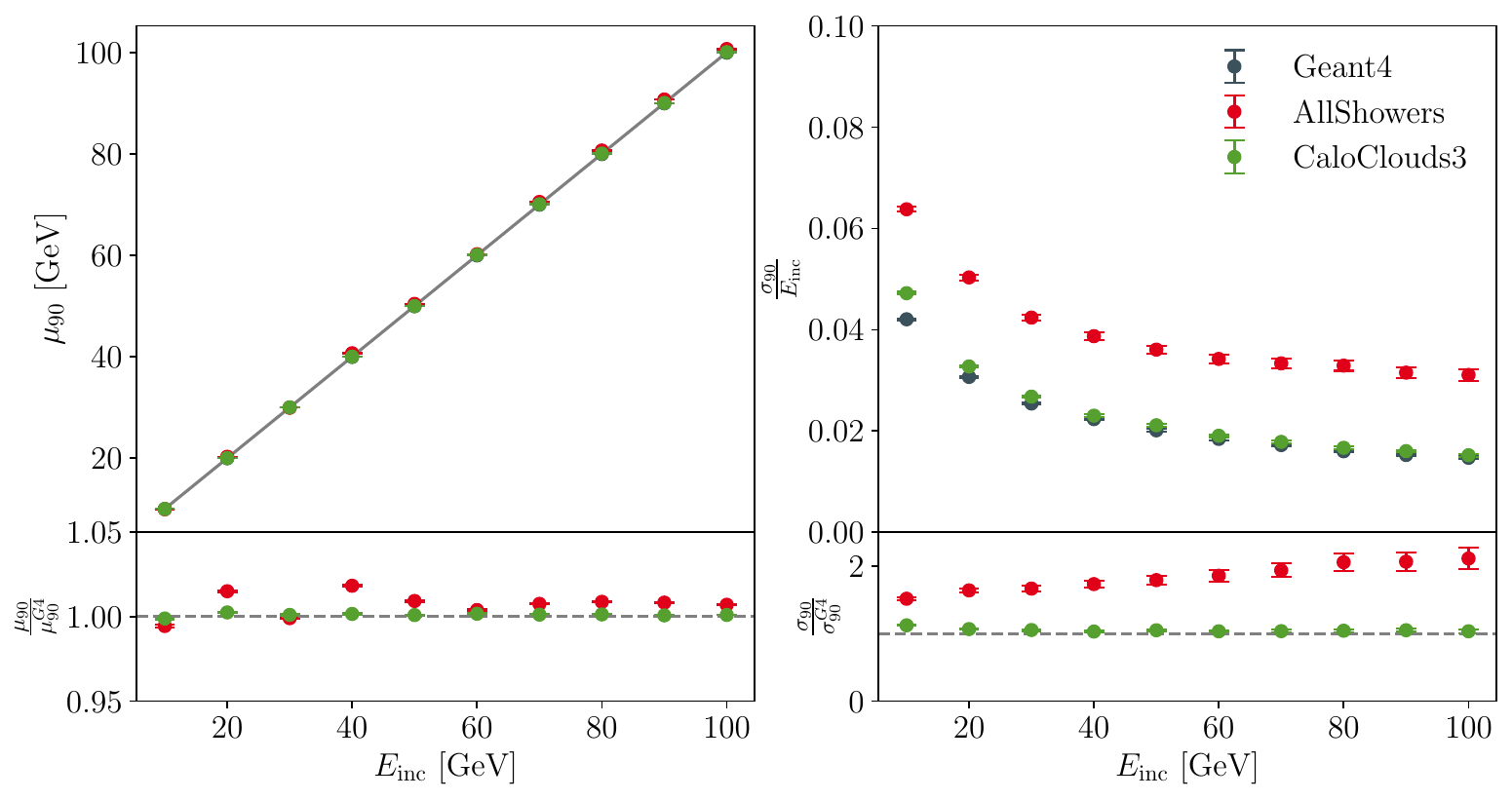}
    \caption{Linearity of rescaled energy sum for photon showers. Incident energies are chosen in steps of 10 GeV between 10 GeV and 100 GeV. Per energy and generator, 10k samples are used. Incident angles are distributed over the intersection of the respective training regions.}
    \label{fig:linearity_photons}
\end{figure}

The linearity of the reconstructed energy is always a key feature for a calorimeter, and must be well replicated in simulations.
In figure~\ref{fig:linearity_photons}, the linearity of photons as simulated by AllShowers and CaloClouds3 is plotted against a Geant4 reference. The simplified energy reconstruction is a linear sum of the energy deposits, with different scaling factors for sections of the calorimeter with different properties.
Three scaling factors are chosen: one for the energy sum of the first 20 ECAL layers, then a second for the energy sum of the last 10 ECAL layers, and finally a factor for the energy sum of the HCAL. All factors are chosen to minimize the mean squared error of the reconstructed Geant4 energies and then applied to both the Geant4 and the two ML model data.

In the reconstructed energy on the left, AllShowers produces agreement with Geant4 on most points; however, some energies show significant deviations. AllShowers does not make energy predictions in PointCountFM; there is only a single energy correction factor applied, see section~\ref{sec:energy_cali}. This correction factor can raise or lower all points collectively, but cannot alter the relative height. By contrast, CaloClouds3 is performing very well across the whole range. Two elements contribute to this: the basic flat profile is achieved by the normalizing flow in CaloClouds3, which predicts energy per layer for the model. Then, in order to obtain the best mean value for all points, a single correction factor is applied in the same way as for AllShowers.

For the energy resolution on the right, the range of reconstructed energies from AllShowers simulations is significantly too wide. This results in higher values (more variance) in the resolution plot. By comparison, CaloClouds3 has a slight deviation in the lower energies, but is otherwise well matched to Geant4. The strong performance here is produced by the dedicated energy per layer predictions made by the normalizing flow in CaloClouds3, which models the variations in energy accurately.

To improve the linearity and energy resolution of AllShowers, it would be possible to add an energy per layer prediction to PointCountFM, in the same manner as is done in CaloClouds3. As AllShowers includes hadronic showers, it is desirable to retain correct energies for points in MIP tracks, and so a simple rescaling of the energy from PointCountFM would be detrimental. It is possible to conceive of various schemes that could rescale the energy per layer, while leaving the energy of MIPs intact, but we leave the exploration of these options to a future work.

\begin{figure}
    \centering
    \begin{subfigure}[T]{0.45\textwidth}
        \includegraphics[width=\textwidth]{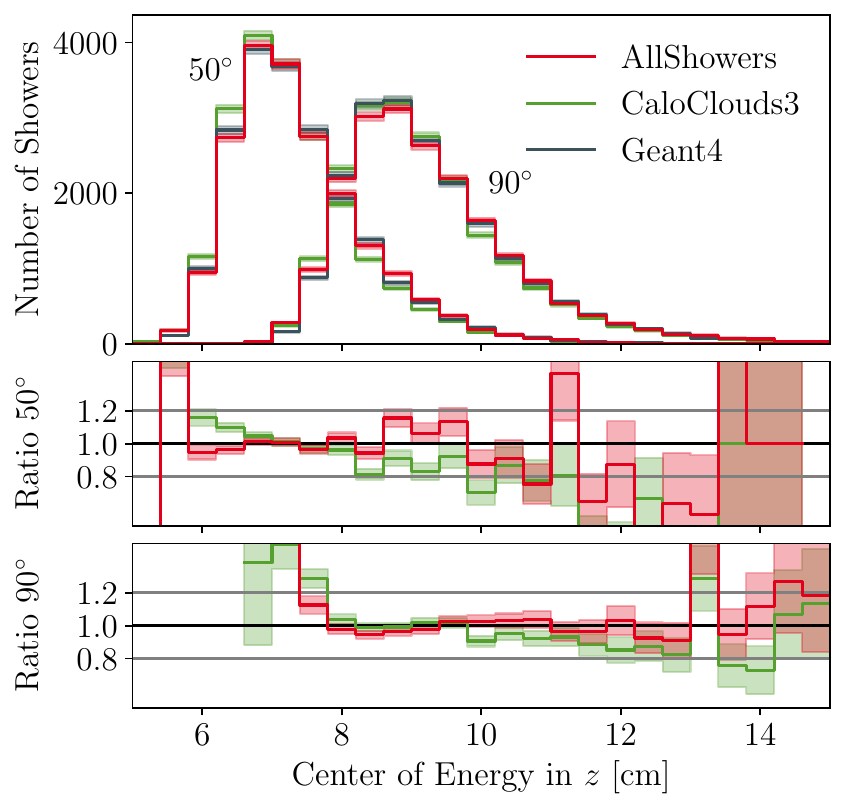}
    \end{subfigure}
    \hspace{1em}
    \begin{subfigure}[T]{0.45\textwidth}
        \includegraphics[width=\textwidth]{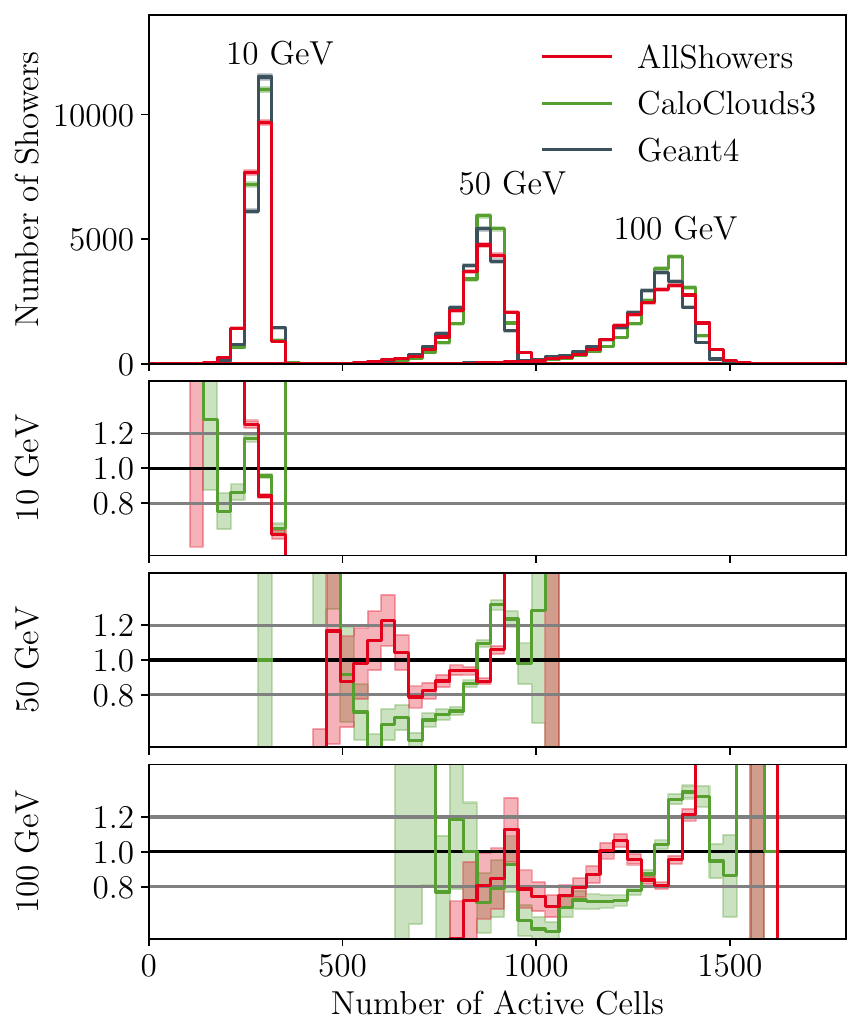}
    \end{subfigure}
    \caption{Center of energy in \(z\)-direction (left) and number of active cells (right) for photon showers fixed incident energy and angle. In the left plot, the incident energy is 50 GeV and in the right plot, the incident angle is \(\theta = 90\degrees\) (perpendicular into the calorimeter). Theta denotes the polar angle in global ILD coordinates. Per generator and incident energy/angle, 20k samples are used.}
    \label{fig:single_angle_energy_photons}
\end{figure}

\autoref{fig:single_angle_energy_photons} shows the center of energy in the \(z\)-direction and the number of active cells for photon showers with fixed incident energy and angle. These plots illustrate how well the model's conditioning on incident energy and angle work. In the upper panels we see that all distributions have been reproduced well by both models. The number of active cells is slightly shifted to lower values for AllShowers. Without calibration of the number of points before clustering, such a shift also would be visible in the CaloClouds3 distribution.

\subsection{Comparison to CaloHadronic}\label{sec:comparison_calo_hadronic}
Another specialized model, which offers a comparison point for \(\pi^+\) showers, is \linebreak[4] CaloHadronic~\cite{Buss:2025cyw}. CaloHadronic is trained only on \(\pi^+\) that enter the calorimeter at a perpendicular angle, so both models will be asked for perpendicular incident angles. The energy range chosen is the full range that CaloHadronic was trained on: \(10\) to \(90\) GeV.

\begin{figure}
    \centering
    \includegraphics[width=0.97\textwidth]{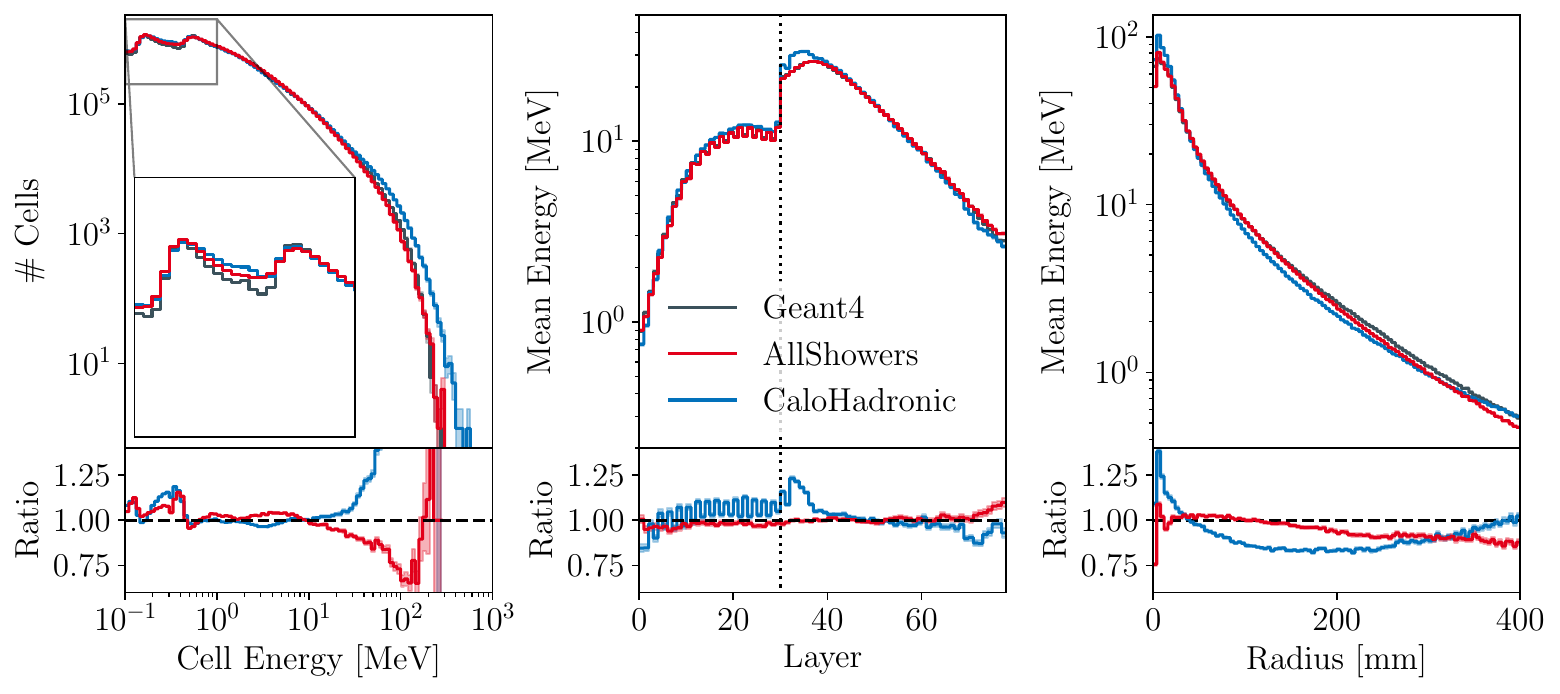}
    \caption{Comparison of AllShowers and CaloHadronic on \(\pi^+\) showers with incident energies uniformly distributed between 10 GeV and 90 GeV. All pions enter the calorimeter perpendicularly. From left to right: cell energy spectrum, longitudinal energy distribution and radial energy distribution. Per generator, 50k samples are used.}
    \label{fig:comparison_calo_hadronic}
\end{figure}

In figure~\ref{fig:comparison_calo_hadronic}, the standard three kinematic profiles are shown for both fast models and Geant4.
On the left, in the cell energy spectrum, both fast models make a reasonably good approximation of the two MIP peaks (one in the ECAL and one in the HCAL).
CaloHadronic significantly overestimates the high-energy tail of the cell energy spectrum, while AllShowers manages to maintain a closer fit to Geant4 for significantly more of the distribution.

In the center, the longitudinal energy distribution for AllShowers is notably better aligned with Geant4 than CaloHadronic.
AllShowers can accurately capture the alternating layer pattern, and also shows better replication of the initial layers of the HCAL. Conditioning on the layer index and allowing to learn the layer properties in an embedding vector (see section~\ref{sec:embeddings}) likely helps here.
Overall, the longitudinal distribution created by AllShowers is remarkably well modeled.

Finally, on the right, we compare the radial distribution for AllShowers and CaloHadronic.
In this distribution, CaloHadronic and AllShowers are more closely matched.
CaloHadronic slightly overestimates the center of the shower, and underestimates a significant portion of the bulk.
AllShowers performs well in the center, with only minor fluctuations in the first few bins,
then tends to underestimate the tail.

\begin{figure}
    \centering
    \includegraphics[width=0.95\textwidth]{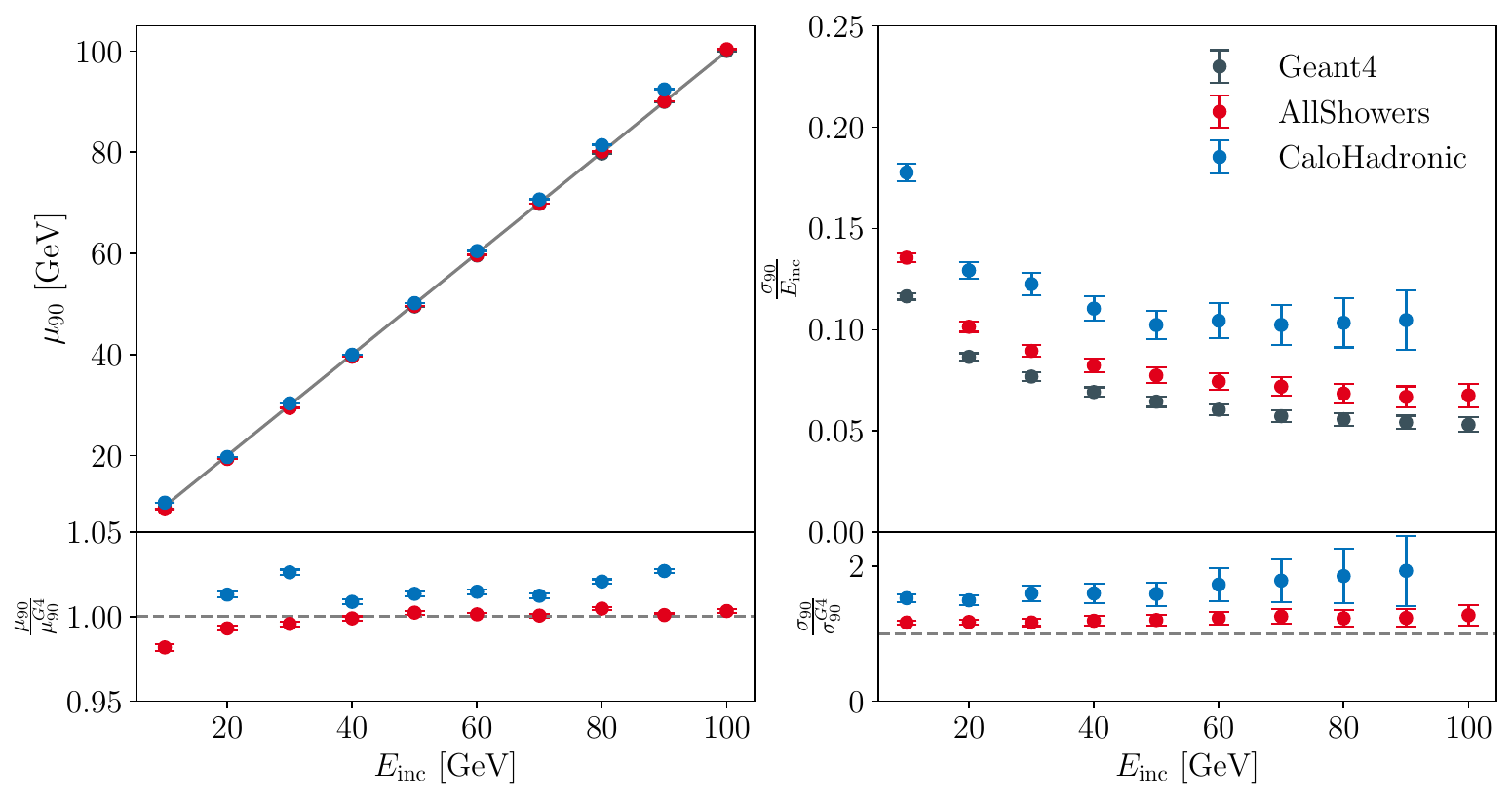}
    \caption{Linearity of rescaled energy sum for \(\pi^+\) showers. Incident energies are chosen in steps of 10 GeV between 10 GeV and 100 GeV. Per energy and generator, 10k samples are used. All pions enter the calorimeter perpendicularly.}
    \label{fig:linearity_pions}
\end{figure}

Reconstructed energy of pions offers important insight into the relationship between energy deposits in the ECAL and HCAL for individual showers.
In figure~\ref{fig:linearity_pions}, we show the linearity and resolution of the reconstructed energy of \(\pi^+\) showers generated by the two fast models and Geant4. The energy reconstruction is performed in the same way as for photons in section~\ref{sec:comparison_calo_clouds}.

AllShowers offers a good reconstructed energy, marginally underestimating the energy of low-energy \(\pi^+\) showers, whereas CaloHadronic consistently overestimates the pion energy.
Looking at the resolution of the reconstructed energies, neither model is performing well.
In both cases, the distribution of the reconstructed energies is too wide at all incident energy points.
AllShowers's performance is better, being within \(50\%\) of Geant4 for all incident energies, but both models leave a lot to be desired in this metric.

Overall, AllShowers clearly provides better kinematic descriptions of \(\pi^+\) showers, with both the performance of the CNF-transformer in the radial direction and the combination of the PointCountFM and the CNF-transformer in the longitudinal direction demonstrating unprecedented accuracy on \(\pi^+\) showers.

\begin{figure}
    \centering
    \begin{subfigure}[T]{0.45\textwidth}
        \includegraphics[width=\textwidth]{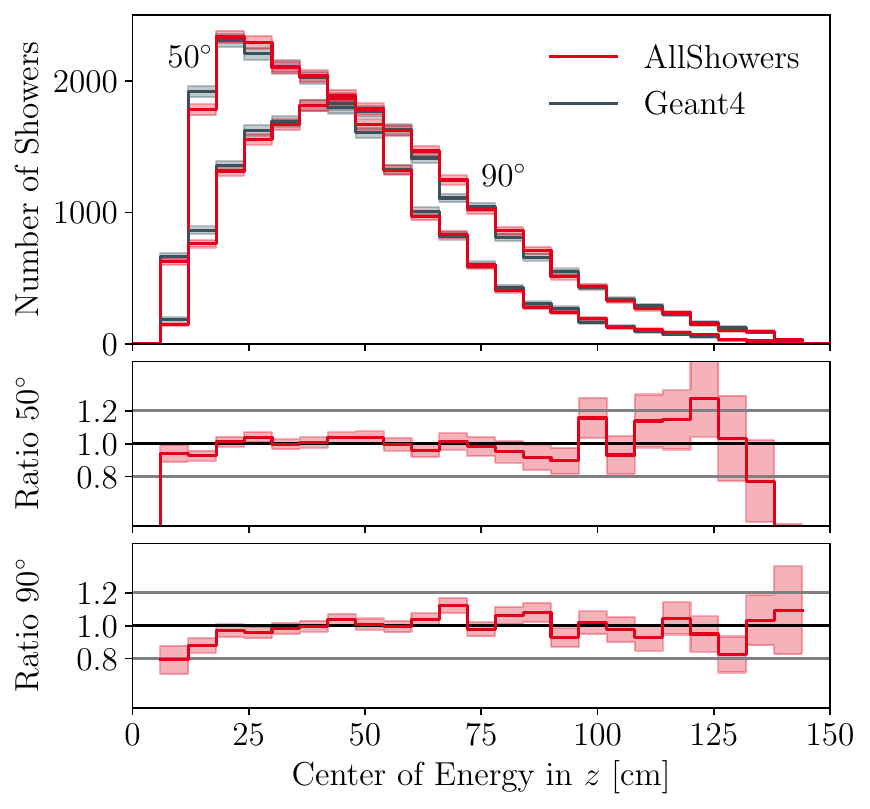}
    \end{subfigure}
    \hspace{1em}
    \begin{subfigure}[T]{0.45\textwidth}
        \includegraphics[width=\textwidth]{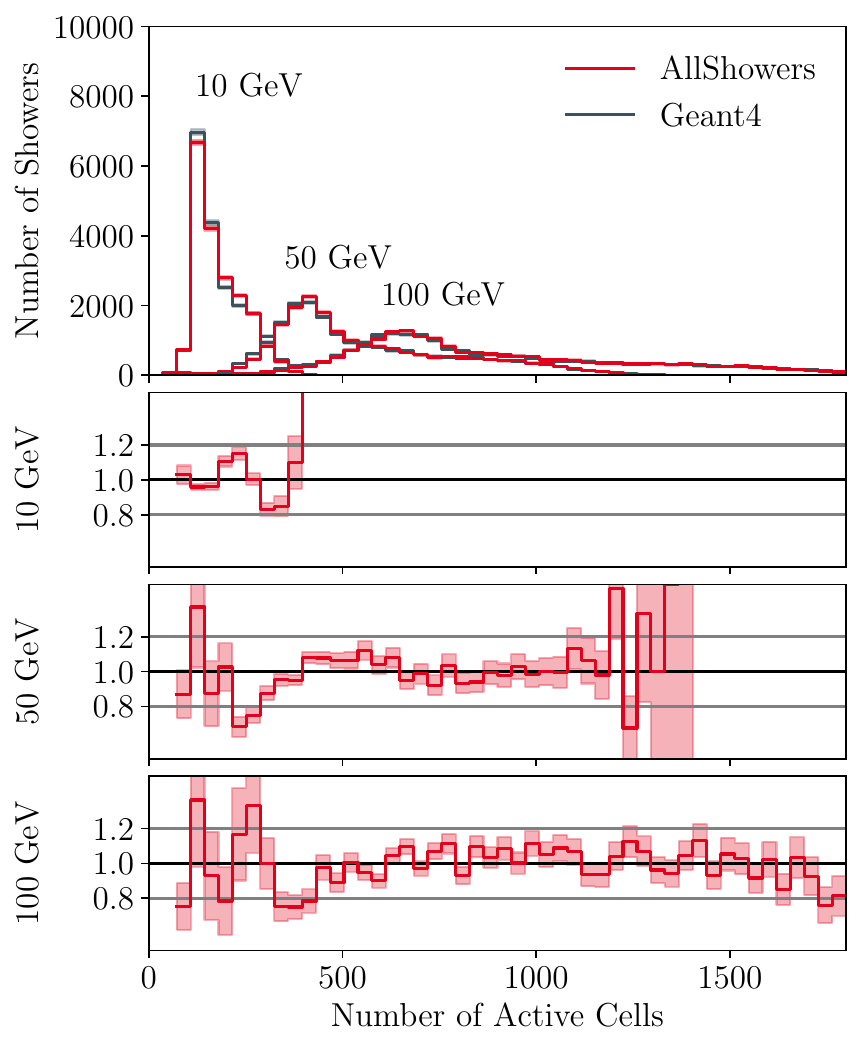}
    \end{subfigure}
    \caption{Center of energy in \(z\)-direction (left) and number of active cells (right) for \(\pi^+\) showers fixed incident energy and angle. In the left plot, the incident energy is 50 GeV and in the right plot, the incident angle is \(\theta = 90\degrees\) (perpendicular into the calorimeter). Theta denotes the polar angle in global ILD coordinates. Per generator and incident energy/angle, 20k samples are used.}
    \label{fig:single_angle_energy_pions}
\end{figure}

\autoref{fig:single_angle_energy_pions} shows the center of energy in the \(z\)-direction and the number of active cells for \(\pi^+\) showers with fixed incident energy and angle. We see good agreement between AllShowers and Geant4 in all distributions. CaloHadronic does not include angular conditioning and the incident energy range is narrower than for AllShowers. Therefore, most of the fixed angle and energy shown here cannot be tested with CaloHadronic.

\subsection{Timing}\label{sec:timing}
When comparing the timing of AllShowers to other models, we provide both the time for the execution of all 32 function evaluations used in the current version of the model and a speculative time needed for a model with only 1 function evaluation.
A reasonable future investigation for AllShowers would be to distill the model.
A distilled model could require as little as a single function evaluation to attain similar performance, but at this point, we have yet to achieve this optimization.
So the timings for 32 function evaluations correspond to the current performance, and the timings for 1 function evaluation are speculative, but provide a good estimate of what timing performance might be attained by the next likely optimization. To get the timing for 1 function evaluation, we generate showers with a single Euler step.

\begin{table}[t]
    \centering
    \begin{tabular}{llrrrr}
        \toprule
        Hardware & Model & NFE & Batch Size & Time / Sample [s] & Speed-factor \\
        \midrule
        CPU & Geant4 & - & 1 & 2.88\phantom{000} & 1.0x \\
        \cmidrule(l){2-6}
        & CaloClouds3 & 1 & 1 & 0.014\phantom{00} & 194.3x \\
        & & & 16 & 0.0041\phantom{0} & 654.5x \\
        \cmidrule(l){2-6}
        & AllShowers & 1 & 1 & 0.17\phantom{000} & 16.7x \\
        \cmidrule(l){3-6}
        & & 32 & 1 & 5.0\phantom{0000} & 0.6x \\
        \midrule
        GPU & CaloClouds3 & 1 & 1 & 0.014\phantom{00} & 208.3x \\
        & & & 16 & 0.00088 & 3256.\phantom{0}x \\
        \cmidrule(l){2-6}
        & AllShowers & 1 & 1 & 0.018\phantom{00} & 164.0x \\
        & & & 16 & 0.0014\phantom{0} & 2114.\phantom{2}x \\
        \cmidrule(l){3-6}
        & & 32 & 1 & 0.054\phantom{00} & 53.2x \\
        & & & 16 & 0.0066\phantom{0} & 440.0x \\
        \bottomrule
    \end{tabular}
    \caption{Timing comparison between Geant4, CaloClouds3, and AllShowers on photon showers.}
    \label{tab:timing_photons}
\end{table}

In table \ref{tab:timing_photons}, we compare the timing of AllShowers to CaloClouds3 and Geant4 on photon showers.
Timings are measured for photons at 9 fixed incident energies, from 10 GeV to 90 GeV in steps of 10 GeV.
The photons are all fired perpendicular to the calorimeter, and 100 batches are simulated per incident energy.
Time of the first two batches is discarded, as the warm-up step may have longer, and more erratic timing, dependent on memory allocation and ``just in time'' compilation.
Times are only for the model's point generation itself; they do not include any overhead for moving data to or from the GPU or projecting hits into the detector geometry.
This slightly complicates the comparison to Geant4, which by design places hits in sensitive cells only and inherently incurs the overhead of the full detector geometry. To allow for a fair comparison, we force our models to use a single computational thread on CPU, as Geant4 does not support parallelism within a single event simulation.

All CPU timings are performed on a single core of an AMD EPYC 7402 processor with 512GB RAM.
All GPU timings are performed on NVIDIA's A100.

CaloClouds3 is a fully distilled model, so 1 function evaluation is all that is ever used.
It has also been aggressively optimized for the specific case of photons, including leveraging photon-specific behaviors, such as the lack of significant substructure in the showers.
With an iid assumption on the points, larger batch sizes become particularly efficient.
AllShowers's current format prohibits specialized treatment of photons,
and being the first generation of this model design, it has not undergone such significant optimization as CaloClouds3, so it is expected that AllShowers cannot compete in inference time with CaloClouds3.
Indeed, CaloClouds3 is at least two orders of magnitude faster on CPU.
On the GPU, the difference is less dramatic, but overall, it is seen that CaloClouds3 will remain significantly faster until AllShowers is distilled or otherwise optimized.

\begin{table}[t]
    \centering
    \begin{tabular}{llrrrr}
        \toprule
        Hardware & Model & NFE & Batch Size & Time / Sample [s] & Speed-factor \\
        \midrule
        CPU & Geant4 & - & 1 & 2.09\phantom{00} & 1.0x \\
        \cmidrule(l){2-6}
        & CaloHadronic & 1 & 1 & 0.59\phantom{00} & 3.5x \\
        & & & 16 & 0.73\phantom{00} & 2.8x \\
        \cmidrule(l){3-6}
        & & 59 & 1 & 34.8\phantom{000} & < 0.1x \\
        & & & 16 & 43.3\phantom{000} & < 0.1x \\
        \cmidrule(l){2-6}
        & AllShowers & 1 & 1 & 0.12\phantom{00} & 16.7x \\
        \cmidrule(l){3-6}
        & & 32 & 1 & 3.5\phantom{000} & 0.6x \\
        \midrule
        GPU & CaloHadronic & 1 & 1 & 0.0086 & 243.0x \\
        & & & 16 & 0.0033 & 633.3x \\
        \cmidrule(l){3-6}
        & & 59 & 1 & 0.40\phantom{00} & 5.3x \\
        & & & 16 & 0.15\phantom{00} & 13.7x \\
        \cmidrule(l){2-6}
        & AllShowers & 1 & 1 & 0.018\phantom{0} & 118.8x \\
        & & & 16 & 0.0014 & 1550.4x \\
        \cmidrule(l){3-6}
        & & 32 & 1 & 0.054\phantom{0} & 38.7x \\
        & & & 16 & 0.0055 & 381.9x \\
        \bottomrule
    \end{tabular}
    \caption{Timing comparison between Geant4, CaloHadronic, and AllShowers on pion showers. Geant4 and CaloHadronic times are taken from \cite{Buss:2025cyw}.}
    \label{tab:timing_pions}
\end{table}

In table~\ref{tab:timing_pions}, a similar timing comparison is shown for pion showers.
This time, CaloHadronic is used as a comparison point, and for CaloHadronic, the NFE is also a tunable parameter. In the table, timings are shown for both a hypothetical distilled version with one function evaluation and for the number of function evaluations used in the current versions of the models.
As with the photon timings, \(\pi^+\) showers are all fired perpendicular to the calorimeter, and 9 fixed energies are simulated between 10 and 100 GeV.
Here, the comparison is closer, as both models are compelled to deal with substructure in hadronic showers.

As CaloHadronic was a pilot model, designed to demonstrate the potential to combine ECAL and HCAL simulation, its code was never restructured to allow compilation.
Thus, if CaloHadronic were timed including preprocessing of input data and postprocessing of generated outputs, it would be unrealistically slow.
Instead, only the evaluation of the PyTorch model itself was timed, as this would dominate the timing in a more realistic deployment.
Due to the omission of all other elements from the timing, the times for CaloHadronic can be regarded as mildly optimistic.
Despite this, AllShowers comes out as faster than CaloHadronic, both at a single function evaluation and with the NFE that is customary for the model.
This shows all-round more efficient use of resources, including good GPU performance.

\begin{figure}
    \centering
    \includegraphics[width=0.48\textwidth]{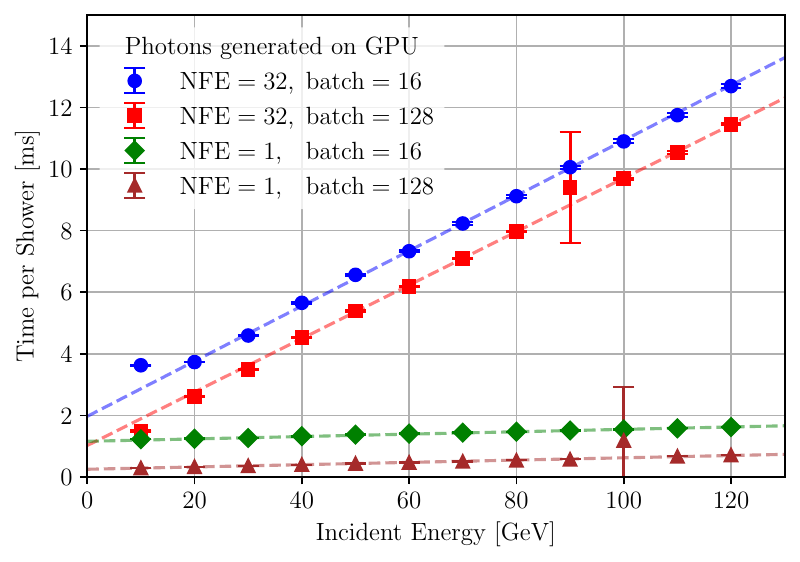}
    \includegraphics[width=0.48\textwidth]{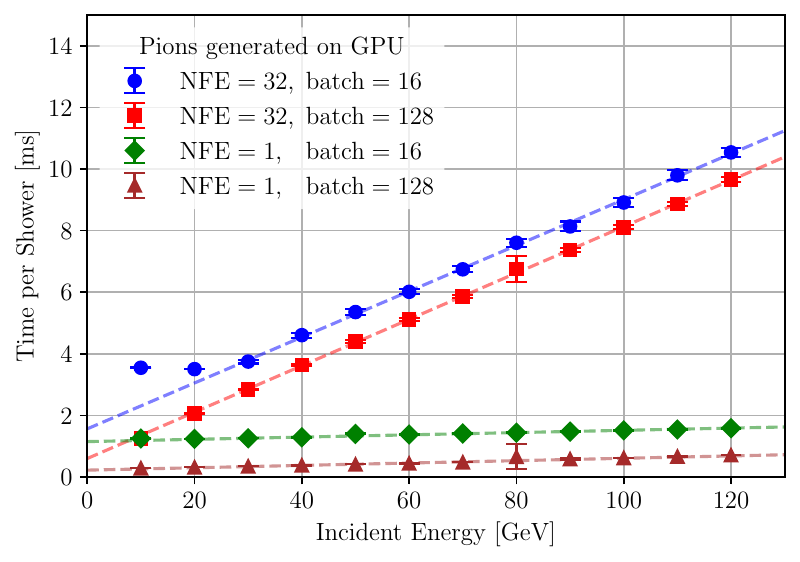}
    \caption{Time per sample generated by AllShowers as a function of incident energy for photons (left) and pions (right). Data points are the mean time per sample for 100 batches. The error bars show the standard deviation of the time per sample across the 100 batches. Dashed lines show linear fits to the data excluding the 10 GeV and 20 GeV points, which in some cases clearly deviate from the linear trend.}
    \label{fig:timing_summary}
\end{figure}
Because the number of energy depositions in a shower is strongly dependent on the incident energy, the sampling time of point-cloud-based generative models is also expected to depend on the incident energy. In figure~\ref{fig:timing_summary}, the time per sample generated by AllShowers is plotted against the incident energy for both photons and pions. A linear fit almost perfectly describes the relationship between generation time and incident energy. We observe positive constant and linear terms in all cases, as expected. Increasing the batch size from 16 to 128 mainly reduces the constant term, while it has little effect on the linear term. This is consistent with the expectation that most overheads are independent of the number of points generated. The scaling of the time per sample with incident energy is a feature of point-cloud-based generative models. Fixed-grid-based generative models do not have this feature, as they generate a fixed number of cells regardless of how many of these cells are activated by the shower~\cite{Buss:2025bec}.

\section{Conclusion}\label{sec:conclusion}
We have presented AllShowers, a novel generative model for high-granularity calorimeter shower simulation. AllShowers is a unified generative model capable of generating multiple particle types, encompassing both electromagnetic and hadronic showers, within a single architecture. This can help reduce the memory footprint, a significant bottleneck in large-scale Monte Carlo production, by allowing loading a single model for all particle types. Moreover, the model is conditioned on incident angle and energy, enabling broad applicability, and it can simultaneously simulate energy depositions across both ECAL and HCAL, thereby enabling end-to-end calorimeter response generation.

AllShowers shows strong agreement with Geant4 across a range of individual-shower features, including aspects of the fine spatial structure accessible with highly granular calorimeters, as well as for ensemble-level distributions spanning multiple particle species. In comparisons at the shower level, its performance is competitive with specialized baselines — \mbox{CaloClouds3} for photons and CaloHadronic for pions — often yielding closer agreement on several observables. For photons, the absence of an iid assumption leads to slower generation than CaloClouds3, while potentially capturing additional correlations in the shower development. However, a definitive assessment of the trade-off between computational performance and physics fidelity for these models ultimately requires evaluating realistic physics observables after full detector reconstruction. For pions, AllShowers achieves comparable or improved agreement relative to CaloHadronic, while also providing faster sampling.

Looking forward, we aim to improve the energy resolution of AllShowers either by generating layer-wise energy deposits similar to CaloClouds3 or by applying a postprocessing step. Additionally, we plan to distill the model to reduce the number of function evaluations (NFE) required at sampling time and to extend AllShowers to additional detector geometries, further broadening its applicability to high-energy physics simulation workflows.

\section*{Acknowledgements}
We would like to thank Martina Mozzanica for providing shower data generated with CaloHadronic for comparison. We thank Anatolii Korol for helping us understand ddsim and the ILC Soft framework, for giving insight into how the training data for CaloHadronic was simulated, and for providing the regularized readout geometry created for earlier datasets. Furthermore, we would like to thank Joschka Birk for fruitful discussions, especially for suggesting that we explore the ranger and lookahead optimizers. We thank Thomas Madlener for providing valuable feedback on the manuscript.

\paragraph*{Funding information}
This research was supported in part by the Maxwell computational resources operated at Deutsches Elektronen-Synchrotron DESY, Hamburg, Germany. This project has received funding from the European Union's Horizon 2020 Research and Innovation programme under Grant Agreement No 101004761. We acknowledge support by the Deutsche Forschungsgemeinschaft under Germany's Excellence Strategy -- EXC 2121 Quantum Universe -- 390833306 and via the KISS consortium (05D23GU4, 13D22CH5) funded by the German Federal Ministry of Research, Technology and Space (BMFTR) in the ErUM-Data action plan.

\begin{appendix}
\numberwithin{equation}{section}

\section{Code and Data Availability}
The code written for this work is available in the following Git repositories:
\begin{itemize}
    \item[] CNF-transformer: \url{https://github.com/FLC-QU-hep/AllShowers}
    \item[] PointCountFM: \url{https://github.com/FLC-QU-hep/PointCountFM}
    \item[] Ranger-Light optimizer: \url{https://github.com/FLC-QU-hep/ranger-lite}
    \item[] Collection of shower IO utilities: \url{https://github.com/FLC-QU-hep/ShowerData}
\end{itemize}
The training datasets simulated for this work are available at:
\begin{itemize}
    \item[] AllShowers Dataset: \url{https://doi.org/10.5281/zenodo.18020348}
\end{itemize}

\section{Number of Trainable Parameters}\label{sec:model_sizes}
\begin{table}[ht]
    \centering
    \begin{tabular}{l r r r}
        \toprule
        Model & Layer-level model & Point-level model & Total \\
        \midrule
        AllShowers & 351,822 & 263,491 & 615,313 \\
        CaloClouds3 & 6,026,520 & 69,640 & 6,096,160 \\
        CaloHadronic & 349,905 & 1,784,724 & 2,134,629 \\
        \bottomrule
    \end{tabular}
    \caption{Number of trainable parameters for AllShowers, CaloClouds3, and CaloHadronic.}
    \label{tab:model_sizes}
\end{table}
In table~\ref{tab:model_sizes}, we compare the number of trainable parameters for AllShowers, CaloClouds3, and CaloHadronic. Shown are the number of parameters in the layer-level model (PointCountFM for AllShowers and CaloHadronic, and the normalizing flow for CaloClouds3), the point-level model (CNF-transformer for AllShowers, and the diffusion models for CaloClouds3 and CaloHadronic), and the total number of parameters. It is evident that AllShowers has a significantly smaller total number of parameters compared to both CaloClouds3 and CaloHadronic.

The number of trainable parameters is not the only factor that influences the memory footprint of a model. A model that uses attention does not necessarily have to store the full attention matrix in memory at once, but substantial parts of it. This can lead to a much higher memory footprint than for models that operate on points independently. To get a good estimate of the memory footprint in a practical application, all models would need to be exported to cpp and integrated into the full simulation chain. With Python interfaces for a batch size of 1, 32 function evaluations, and computations on a single CPU thread, we find that AllShowers uses up to 1.8 GB of RAM during setup and up to 0.9 GB of RAM during sampling.

\section{Hyperparameters}\label{app:hyperparameters}
\begin{table}[p]
    \begin{tabular}{l l r}
        \toprule
        \textbf{Type} & \textbf{Parameter} & \textbf{Value} \\
        \midrule
        Data Preprocessing & incident particle type & one-hot encoding \\
        & incident energy & Standard Scaling of \(E_{\text{inc}}\) \\
        & incident angle & unit sphere representation \\
        & point per layer & Standard Scaling of \(\log(0.5 + N_{i})\) \\
        \midrule
        Model & hidden layers & 5 \\
        & hidden dims & 128, 256, 512, 256, 128 \\
        & activation & ReLU \\
        \midrule
        Training & optimizer & Adam \\
        & learning rate scheduler & OneCycle \\
        & maximum learning rate & \(10^{-3}\) \\
        & batch size & 1024 \\
        & epochs & 1000 \\
        \midrule
        Sampling & ODE solver & Heun \\
        & NFE & 100 \\
        \bottomrule
    \end{tabular}
    \caption{Hyperparameters used for the PointCountFM model.}
    \label{tab:PointCountFM-parameters}
\end{table}
\begin{table}[p]
    \begin{tabular}{l l r}
        \toprule
        \textbf{Type} & \textbf{Parameter} & \textbf{Value} \\
        \midrule
        Data Preprocessing & point \(x, y\) & Standard Scaling \\
        & point energy & Standard Scaling of \(\log(E)\) \\
        & incident energy & Standard Scaling of \(\log(E_{\text{inc}})\) \\
        & flow time & Fourier embedding with 3 frequencies \\
        & incident angle & unit sphere representation \\
        & OT mapping & layer-and-shower-wise \\
        \midrule
        Model & embedding dim & 64 \\
        & transformer encoder blocks & 4 \\
        & attention heads & 4 \\
        & feedforward dim & 256 \\
        & activation & GELU \\
        & attention masking & custom calorimeter-layer-based \\
        \midrule
        Training & optimizer & Ranger (Lookahead + RAdam) \\
        & learning rate scheduler & cosine annealing \\
        & initial learning rate & \(10^{-3}\) \\
        & weight decay & \(10^{-2}\) \\
        & gradient clipping & 0.2 \\
        & batch size & 256 \\
        & epochs & 200 \\
        \midrule
        Sampling & ODE solver & midpoint \\
        & NFE & 32 \\
        \bottomrule
    \end{tabular}
    \caption{Hyperparameters used for the CNF-transformer model.}
    \label{tab:CNF-Transformer-parameters}
\end{table}
All hyperparameters used to train PointCountFM can be found in table~\ref{tab:PointCountFM-parameters} and those used to train the CNF-transformer in table~\ref{tab:CNF-Transformer-parameters}.

\section{Additional Plots for Multiple Particle Types}\label{app:additional_plots}
\begin{figure}[p]
    \centering
    \includegraphics[width=0.96\textwidth]{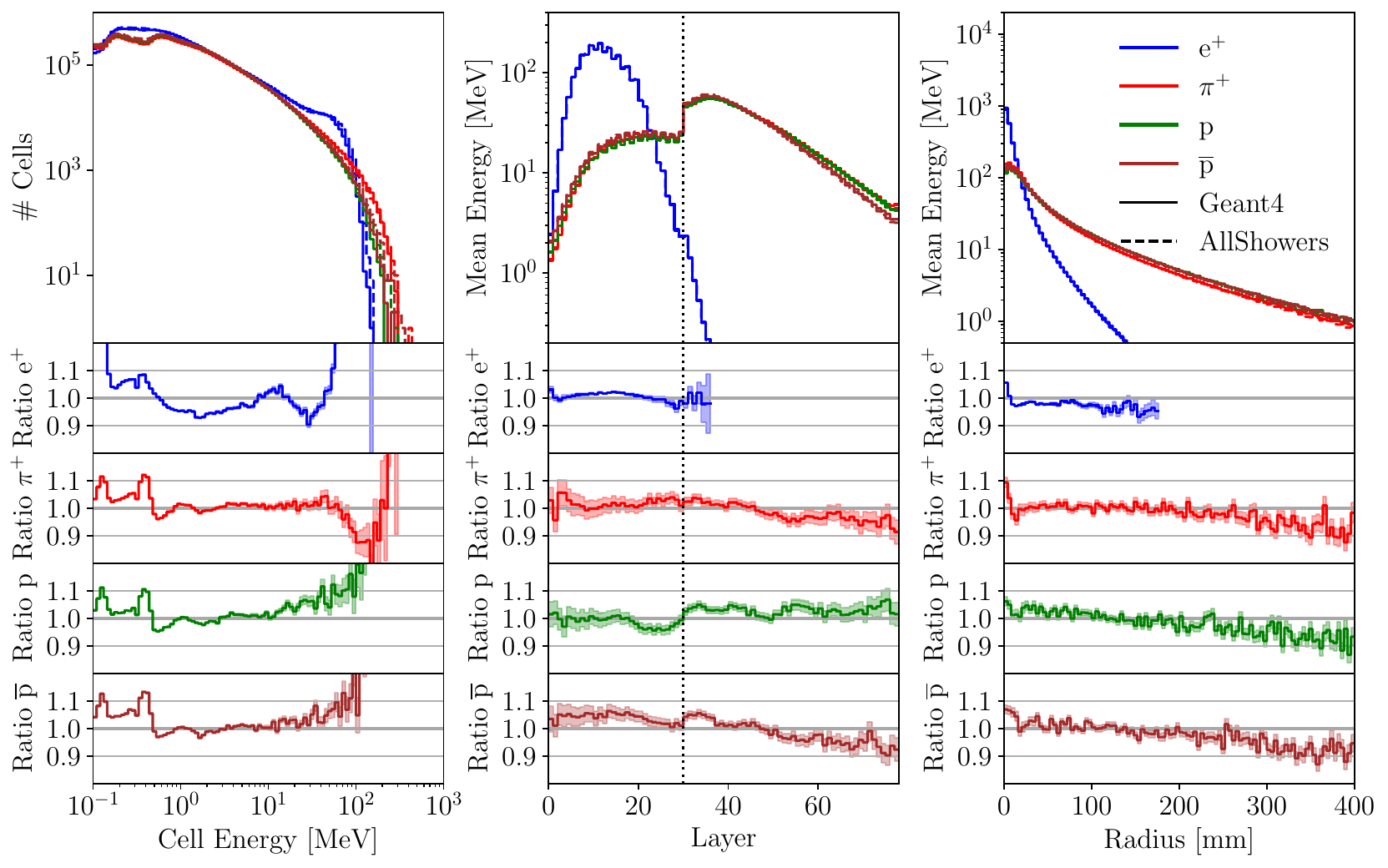}
    \includegraphics[width=0.96\textwidth]{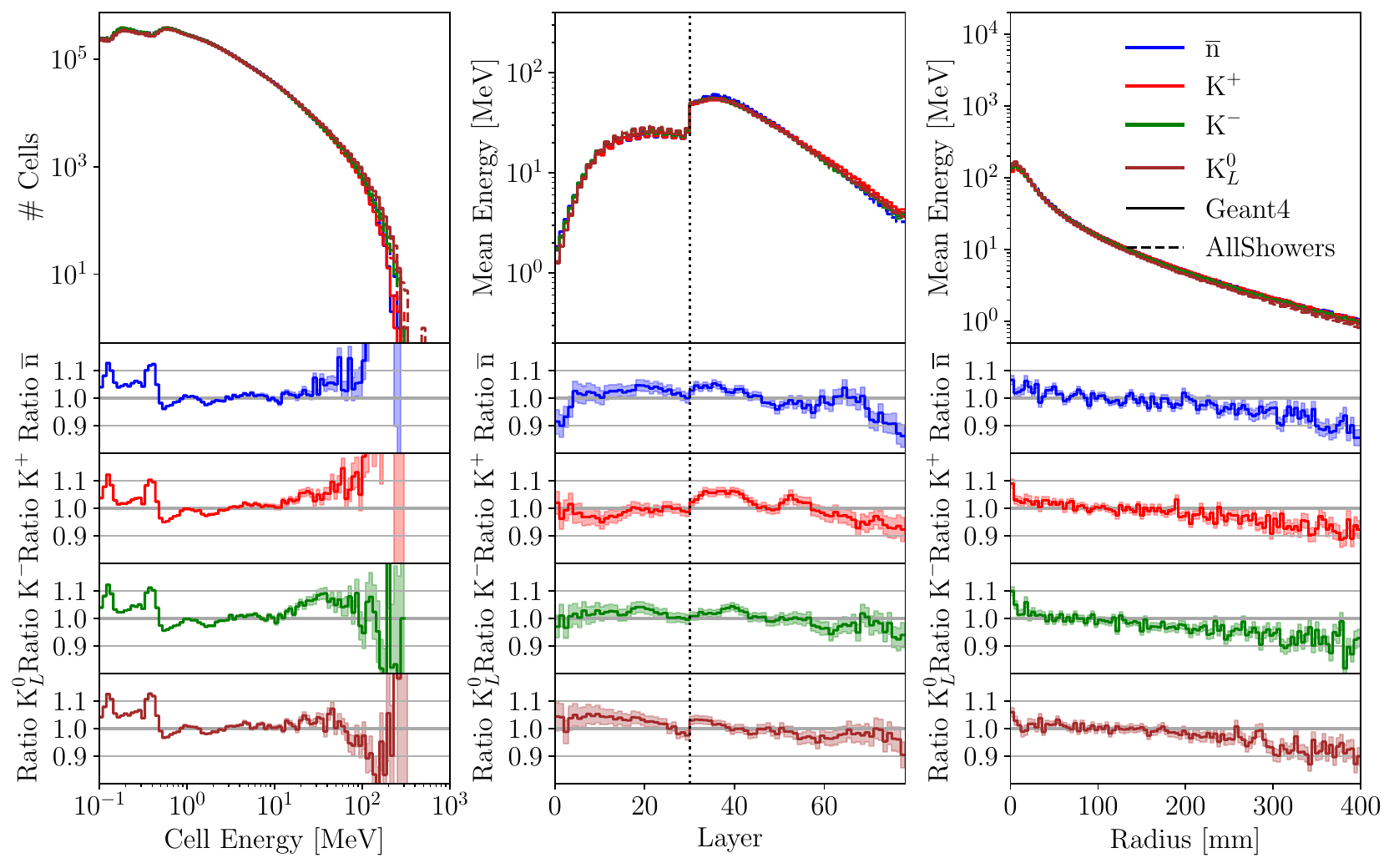}
    \caption{Analogous plot to figure~\ref{fig:distributions_all_types}, showing the incident particle types left out of the main text.}
    \label{fig:multi_particle_types_2}
\end{figure}
\autoref{fig:multi_particle_types_2} shows the same distributions as in \autoref{fig:distributions_all_types}, but for the incident particle types left out of the main text. The plots confirm that the performance of AllShowers is consistent across all particle types it was trained on.

\section{Angular Reconstruction}\label{app:angular_reconstruction}
The internal angle reconstruction we use for \autoref{fig:distributions_all_types} is based on a principal component analysis (PCA) of the hit positions with energy weighting. In a first step, the PCA is performed on all hits in the shower. The first principal component is then used to define the shower axis. When the reconstructed direction points away from the calorimeter, the direction is flipped. In a second step, all hits outside a cylinder around the reconstructed shower axis are removed, and the PCA is performed again on the remaining hits. This process is repeated three times after which it has converged to a stable solution. Because of the quadratic term in the PCA, outliers can have a significant impact and removing them improves the reconstruction. The radius of the cylinder is set to 16 mm for photons and 80 mm for neutral hadrons. These values are chosen to minimize the angular resolution error for the Geant4 dataset.

\end{appendix}

\bibliography{bibliography}

\end{document}